\documentclass[pre,aps,twocolumn, floatfix, superscriptaddress,showpacs]{revtex4-1}
\usepackage{graphicx}
\usepackage{subfigure}
\usepackage{amsmath}
\usepackage{amsfonts}
\usepackage{amssymb}
\usepackage{float}
\newcommand{\dd}[2]{\frac{d #1}{d #2}}
\newcommand{\pd}[2]{\frac{\partial #1}{\partial #2}}
\newcommand{\pdd}[2]{\frac{\partial^2 #1}{\partial #2^2}}

\begin{document}
\title{Fitness voter model: damped oscillations and anomalous consensus}
\author{Anthony Woolcock}
\affiliation{Centre for Complexity Science, University of Warwick, Gibbet Hill Road, Coventry CV4 7AL, UK}
\email{A.J.Woolcock@warwick.ac.uk}
\author{Colm Connaughton}
\email{connaughtonc@gmail.com}
\affiliation{Mathematics Institute, University of Warwick, Gibbet Hill Road, Coventry CV4 7AL, UK}
\affiliation{Centre for Complexity Science, University of Warwick, Gibbet Hill Road, Coventry CV4 7AL, UK}
\affiliation{London Mathematical Laboratory, 14 Buckingham St., London WC2N 6DF, UK}
\affiliation{Kavli Institute for Theoretical Physics, University of California, Santa Barbara, CA 93106, USA}
\author{Yasmin Merali}
\affiliation{Hull University Business School, University of Hull, Hull, UK, HU6 7RX}
\email{y.merali@hull.ac.uk}
\author{Federico Vazquez}
\email{fede.vazmin@gmail.com}
\affiliation{IFLYSIB, Instituto de F\'{i}sica de L\'{i}quidos y Sistemas Biol\'{o}gicos (UNLP-CONICET), 1900 La Plata, Argentina}
\affiliation{Centre for Complexity Science, University of Warwick, Gibbet Hill Road, Coventry CV4 7AL, UK}
\date{\today}

\begin{abstract}
We study the dynamics of opinion formation in a heterogeneous voter
model on a complete graph, in which each agent is endowed with an
integer fitness parameter $k \ge 0$, in addition to its $+$ or $-$
opinion state.  The evolution of the distribution of $k$--values and
the opinion dynamics are coupled together, so as to allow the system
to dynamically develop heterogeneity and memory in a simple way.  When
two agents with different opinions interact, their $k$--values are
compared and, with probability $p$ the agent with the lower value
adopts the opinion of the one with the higher value, while with
probability $1-p$ the opposite happens.  The winning agent then
increments its $k$--value by one.  We study the dynamics of the system
in the entire $0 \le p \le 1$ range and compare with the case $p=1/2$,
in which opinions are decoupled from the $k$--values and the dynamics
is equivalent to that of the standard voter model.  When $0 \le p <
1/2$, agents with higher $k$--values are less persuasive, and the
system approaches exponentially fast to the consensus state of the
initial majority opinion.  The mean consensus time $\tau$ appears to
grow logarithmically with the number of agents $N$, and it is greatly
decreased relative to the linear behavior $\tau \sim N$ found in the
standard voter model.  When $1/2 < p \le 1$, agents with higher
$k$--values are more persuasive, and the system initially relaxes to a
state with an even coexistence of opinions, but eventually reaches
consensus by finite-size fluctuations.  The approach to the
coexistence state is monotonic for $1/2 < p < p_o \simeq 0.8$, 
while for $p_o \le p \le 1$ there are damped oscillations around the
coexistence value.  The final approach to coexistence is approximately
a power law $t^{-b(p)}$ in both regimes, where the exponent $b$ increases with 
$p$.  Also, $\tau$ increases respect to the standard voter model,
although it still scales linearly with $N$.  The $p=1$ case is
special, with a relaxation to coexistence that scales as $t^{-2.73}$
and a consensus time that scales as $\tau \sim N^\beta$, with $\beta
\simeq 1.45$.   
\end{abstract}

\maketitle

\section{Introduction and motivation}
\label{sec-intro}

Simple agent-based models of social interactions have attracted a lot
of interest from statistical physicists in recent times. See the
review \cite{castellano_statistical_2009} and the references
therein. From the social sciences perspective such models can provide
a controlled testing ground for more qualitative theories of social
interaction. See, for example, the discussion
in \cite{macy_factors_2002}. For physicists, these  models are
intellectually stimulating because they exhibit a rich range of
dynamical and statistical phenomena. Furthermore, as quantitative data
become increasingly available, particularly from online social
networks \cite{ciulla_beating_2012} and search engine
data \cite{preis_complex_2010}, there is the possibility that social
modelling could become predictive \cite{conte_manifesto_2012}. 

The statistics of consensus formation in opinion dynamics models is a
problem which has proven to be particularly amenable to analysis using
tools from statistical physics, such as the theory of coarsening and
first-passage phenomena (see \cite[chap.~8]{krapivsky_kinetic_2010}).
The primary question of interest is whether the population will
eventually reach consensus or a coexistence of both opinions will
persist indefinitely.  In principle, both outcomes are possible
depending on the assumptions made about the mechanism of social
influence \cite{flache_local_2011}.  The voter
model \cite{clifford_model_1973,ligget_1975,liggett_interacting_2004}
is the most basic opinion dynamics model that allows to explore the
consensus problem in great detail.  It is a simple interacting
particle system consisting of a graph with an agent at each node
possessing a single degree of freedom, its opinion $s$, taking two
possible values.  The dynamics is as follows: an agent is picked at
random who then adopts the opinion of a randomly selected
neighbour. Social influence is therefore represented as an entirely
mindless process whereby agents just adopt the opinions of their
neighbours at random.   The model has different interpretations aside
from opinion dynamics, including
catalysis \cite{krapivsky_kinetics_1992} and population
dynamics \cite{clifford_model_1973}.  Also, different extensions of
the model have been proposed in the literature, including constrained
interactions \cite{Vazquez_2003,Vazquez_2004}, non-equivalent
states \cite{Castello_2006}, asymmetric transitions or
bias \cite{Antal_2006}, noise \cite{Medeiros_2006}, and memory
effects \cite{DallAsta_2007}. It is also known that several models
presenting a coarsening process without surface tension belong to the
voter model universality
class \cite{Dornic_2001,AlHammal_2005,Vazquez_2008b}.

On a finite graph, the voter model always reaches consensus by
finite-size fluctuations, because the two states in which all opinions
are the same are absorbing. This means that once the system randomly
enters such a configuration, it stays there forever.  The average time
it takes a system containing $N$ agents to reach consensus is called
the mean  consensus time, $\tau$, which grows as $\tau \sim N^2$ on a
regular $1D$ lattice, as $\tau \sim N\,\ln N$ on a regular $2D$
lattice, and as  $\tau \sim N$ in $3D$ and above, including on the
complete graph (in which every agent is connected to every other).

It is natural to ask how robust are the results described
above. Research has shown that changes to the voter model, which at
first sight  might seem innocuous, can lead to significant changes in
the statistics of consensus times. In particular, the introduction of
various forms of  heterogeneity can have far--reaching
consequences. Heterogeneity here means that not all agents are
equivalent.  The introduction of even a single
``zealot'' \cite{mobilia_does_2003} -- an agent with a  finite
probability to unilaterally change opinion back to a preset preference
-- greatly increases the consensus time. Also, if all agents are
assigned flip rates which are sampled from a probability distribution
then, by choosing this distribution  appropriately, one can make the
approach to consensus arbitrarily
slow \cite{masuda_heterogeneous_2010}. This model is called the
Heterogeneous Voter Model. On the complete graph, one  can have
$\tau \sim  N^\beta$ with $\beta>1$ arbitrarily large. This occurs
because $\tau$ is dominated by the (slowest) flip rates of the
``stubbornest'' agents. A related model is the Partisan Voter
Model \cite{masuda_heterogeneous_2010}, in which heterogeneity is
introduced by randomly endowing all agents with a  preferred
opinion. The interaction rules are modified so that agents have a
higher rate for switching to their preferred opinion. The consensus
time on  the complete graph was then found to be even longer and grow
exponentially with $N$. In contrast to these examples, heterogeneity
can also accelerate  the formation of consensus. For example, if the
agents acquire temporal memory such that their flip rates decrease the
longer they remain in a given  state then, depending on the strength
of this effect, the scaling exponent of $\tau$ with $N$ decreases
relative to standard voter model
case \cite{stark_decelerating_2008,stark_slower_2008}. Even more
extremely, if the underlying graph is replaced by a random, scale-free
network, then depending on the properties of the degree distribution
of this network, $\tau$ can scale sub-linearly with $N$ and can even
become logarithmic or independent of
$N$ \cite{sood_voter_2005,Suchecki_2005,Vazquez_2008a}. This effect is
known as ``fast consensus''.  Finally, in reference \cite{Perez_2016}
the authors studied a two-sate interacting particle model with
age-dependent transition rates, that is, where the likelyhood that a
given particle changes state depends on the time spent on its actual
state.  This model exhibits either dominance of old states or
coexistence and synchronized collective behavior, depending on whether
particles have a preference to adopt old or new states, respectively.  

In this paper, we study a variant of the voter model in which, unlike
the examples given above, heterogeneity is allowed to develop
dynamically as a  result of the interactions between agents. There are
many ways in which one could imagine doing this. Our approach is one
of the simplest in the sense  that we have a single control
parameter. Each agent is endowed with a fitness parameter, $k$, in
addition to its opinion.  Initially this parameter is  set to zero for
all agents. When two agents with different opinions interact,
$k$--values are compared.  Then, with probability $p$ the agent with
the lower value adopts the  opinion of the one with the higher value
and, with the complementary probability $1-p$ the opposite
happens. The winning agent then increments its $k$--value by one. This
competitive aspect was motivated by the study of the evolution of
competitive societies presented in \cite{ben-naim_structure_2006}. The
only parameter in the model is the  probability $p$. As time passes,
heterogeneity develops as a distribution of different $k$-values
evolves in the population. The opinion group exchange dynamics  has memory
since the transition rates depend on the past history through the
$k$-values, although the model can obviously be formulated in a
Markovian  way in the extended $(s,k)$ state space. We find that the
dynamics of this model is surprisingly rich as the parameter $p$ is
varied.  In particular, it can exhibit both fast and slow consensus
for different values of $p$, and the dynamics of the group sizes
exhibits interesting oscillations in time.

The layout of the paper is as follows. In Sec.~\ref{sec-model} we
specify the model and explain some of the basic properties.
Sec.~\ref{sec-consensusTimes} then presents some numerical
measurements of the consensus time on a complete graph as a function
of $p$ and $N$.  We find that for $p>1/2$ the time to reach consensus
is very large, whereas for $p<1/2$ consensus is very fast.  In
Sec.~\ref{sec-MF}  we study the rate equations for the model and
discover that when $p>1/2$ the system is attracted to a coexistence
state in which two equal sized populations of oppositely--opinioned
agents reach a dynamic equilibrium.  The approach to this state could
be either monotonic or oscillatory.  When $p<1/2$, the coexistence
state is unstable and the system is driven quickly to consensus.  
We finish in
Sec.~\ref{sec-conclusions} with a short summary and conclusions.

\section{Definition of the model}
\label{sec-model}

Two groups, labeled ($\pm$), compete for membership  in a population
of $N$ agents. The interaction network is a complete graph so any
agent can interact with any other. In addition to its group
designation, or opinion $s=\pm$, each agent has an integer fitness
$k$.  In a single time step $\Delta t=2/N$, two agents are selected at 
random.  If both have the same opinion nothing happens. If
not, they interact as follows. Agents compare their respective
$k$--values. With probability $p$, the agent with the lower $k$--value
adopts the opinion of the agent with the higher value. With
probability $1-p$, the opposite happens and the agent with the higher
$k$--value adopts the opinion of the agent with the lower value. If
both $k$--values are equal then one agent adopts the opinion of the
other with equal probability $1/2$. The ``winning'' agent (i.e. the
one whose opinion remained unchanged) then increments its $k$--value
by one. Assuming that the chosen agents have opinions $s_1$ and $s_2$
(with $s_1 \neq s_2$), and fitness $k_1$ and $k_2$, the interaction rules
can be summarized schematically as:

If $k_1>k_2$ then
\begin{alignat}{4}
\nonumber (s_1,k_1)\oplus (s_2,k_2) &\rightarrow&&
\left\{
\begin{array}{ll} 
(s_1,k_1+1)\oplus (s_1,k_2) &\mbox{Prob. $p$} \\ (s_2,k_1)\oplus
(s_2,k_2+1) &\mbox{Prob. $1-p$}.
\end{array}
\right.
\end{alignat}

If $k_1=k_2=k$ then
\begin{alignat}{4}
\nonumber (s_1,k)\oplus (s_2,k) &\rightarrow&&
\left\{
\begin{array}{ll} 
(s_1,k+1)\oplus (s_1,k) &\hspace{0.31cm}\mbox{Prob. $1/2$} \\
(s_2,k)\oplus (s_2,k+1) &\hspace{0.31cm}\mbox{Prob. $1/2$}.
\end{array}
\right.
\end{alignat}
The only parameter in this model is the probability $p$. When $p=1/2$,
the $k$--values of the agents play no role in the opinion dynamics
since each agent has a 50--50 chance of adopting the opinion of the
other. Thus for $p=1/2$, the evolution of the $k$--values decouple
from the opinion dynamics which is therefore equivalent to the
standard voter model. If $p>1/2$, a higher value of $k$ makes an agent
less likely to change opinion during an interaction. Thus, in this
regime, one could interpret the above rules as saying that agents
become more confident of their opinion each time they succeed in
convincing other agents to switch groups. As $p$ gets close to one,
agents which reach high $k$--values become highly unlikely to change
their opinion. Thus ``zealots'' emerge dynamically in this model. If
$p<1/2$, the opposite is true: agents with high $k$--values are more
likely to change opinion frequently.

\section{Consensus times}
\label{sec-consensusTimes}

\begin{figure}
\begin{center}
\includegraphics[width=\columnwidth]{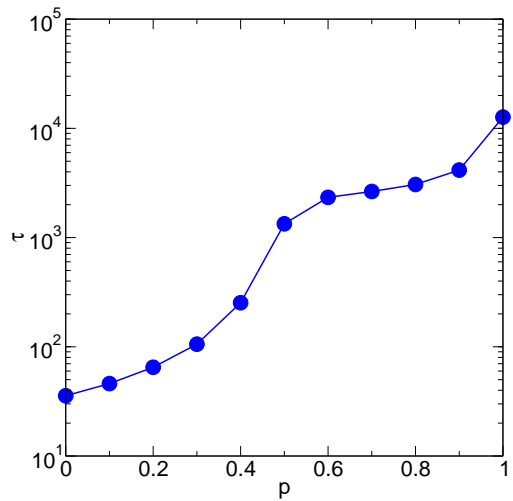}
\caption{Linear-log plot of the mean consensus time $\tau$ vs the probability
  $p$ that the fitter agent wins in an interaction, for a system of
  $N=1000$ agents.} 
\label{fig-Tau-p}
\end{center}
\end{figure}

\begin{figure}
\begin{center}
\includegraphics[width=\columnwidth]{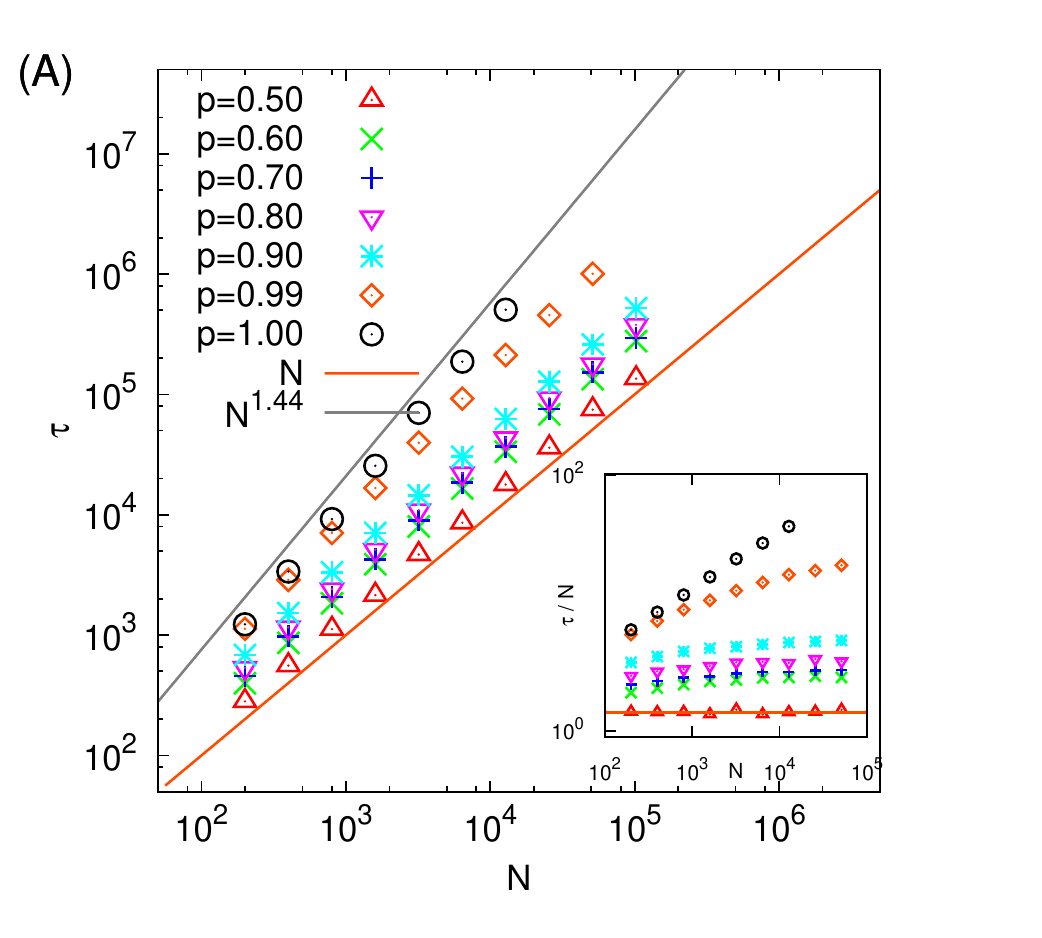}\\ 
\includegraphics[width=\columnwidth]{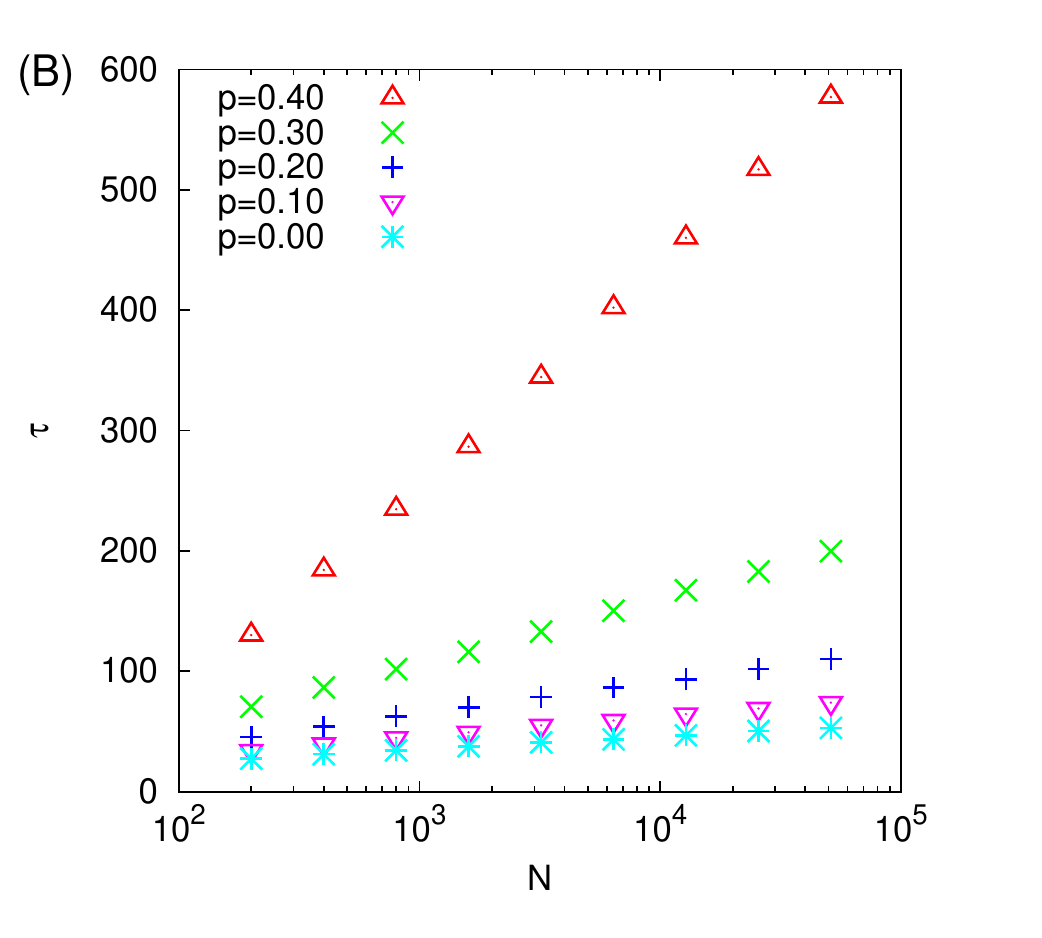}
\caption{(A) Mean consensus time $\tau$ as a function of the number of 
  agents $N$ on a log-log scale, for $p$ in the range $\frac{1}{2}
  \leq p \leq 1$.  Upper and lower straight lines have slope $1.45$
  and $1.0$, respectively.  Inset: $\tau$ is compensated by $N$. The
  horizontal line is the prefactor $2 \ln(2)$ of the standard voter
  model ($p=1/2$).  (B) $\tau$ vs $N$ on a log-linear scale for  $0 \leq
  p < \frac{1}{2}$. }
\label{fig-consensusTimes}
\end{center}
\end{figure}

We performed Monte Carlo simulations of the dynamics defined in
Sec.~\ref{sec-model} above on complete graphs of various sizes $N$ and
for several values of $p$ in the interval $[0,1]$, and measured the
time to reach opinion  consensus.  Initially, each agent takes
opinion $+$ or $-$ with the same probability $1/2$, and all agents
have fitness $k=0$.  Results are shown in Figs.~\ref{fig-Tau-p} and
\ref{fig-consensusTimes}.  As discussed above, the basis for
comparison is the $p=1/2$ case where the opinion dynamics is
equivalent to the  standard voter model for which the consensus time
scales as $\tau = 2 \ln(2) N$ on the complete graph, when the initial
state consists on $N/2$ agents in each opinion state.  

In Fig.~\ref{fig-Tau-p} we plot the mean consensus time $\tau$
measured from simulations, as a function of the probability $p$.  The
average was done over $10^4$ independent realizations of the dynamics.
We observe that $\tau$ increases rapidly with $p$, and becomes very
large when $p$ overcomes the value $1/2$.  This is an indication that
the behaviour for 
$p>1/2$ is strikingly different from that of $p<1/2$.  
Figure~\ref{fig-consensusTimes}(A) shows $\tau$ as a function of $N$ for a range
of values of $p$ between $1/2$ and $1$.  With the exception of the
curve for $p=1$, all curves grow as $N$ for large $N$ (lower straight
line).  This can be seen in the inset of
Fig.~\ref{fig-consensusTimes}(A), where $\tau$ is 
compensated by $N$.  All curves for $p<1$ reach a plateau as $N$
grows, although the saturation level increases with $p$. Thus, the
consensus time grows with $p$ although it still scales linearly with
$N$ as in the standard voter model.  The case $p=1$ seems
quantitatively different.  A closer analysis of the data indicates
that for $p=1$, $\tau  \sim N^\beta$ with $\beta \simeq
1.45$ (upper straight line). Simulations with values of $p$ very close to
$1$ (see $p=0.99$ curve) suggest that the case $p=1$ really is
uniquely different.  As best as we have been able to tell from the
numerics, the scaling exponent jumps at $p=1$.  As we see in
Fig.~\ref{fig-consensusTimes}(B), the behaviour of $\tau$ for $p<1/2$
is quite different.  We observe that $\tau$ grows very slowly with the
system size, as $\tau \sim \ln N$, indicating that the system goes to
a ``fast consensus''. 

In summary, the consensus is slow for $p \ge 1/2$ and very fast for
$p<1/2$.  In an attempt to get some insight into these behaviours, we
shall now study the rate equations corresponding to this model.

\section{Mean field dynamics}
\label{sec-MF}

\begin{figure}
\begin{center}
\includegraphics[width=\columnwidth]{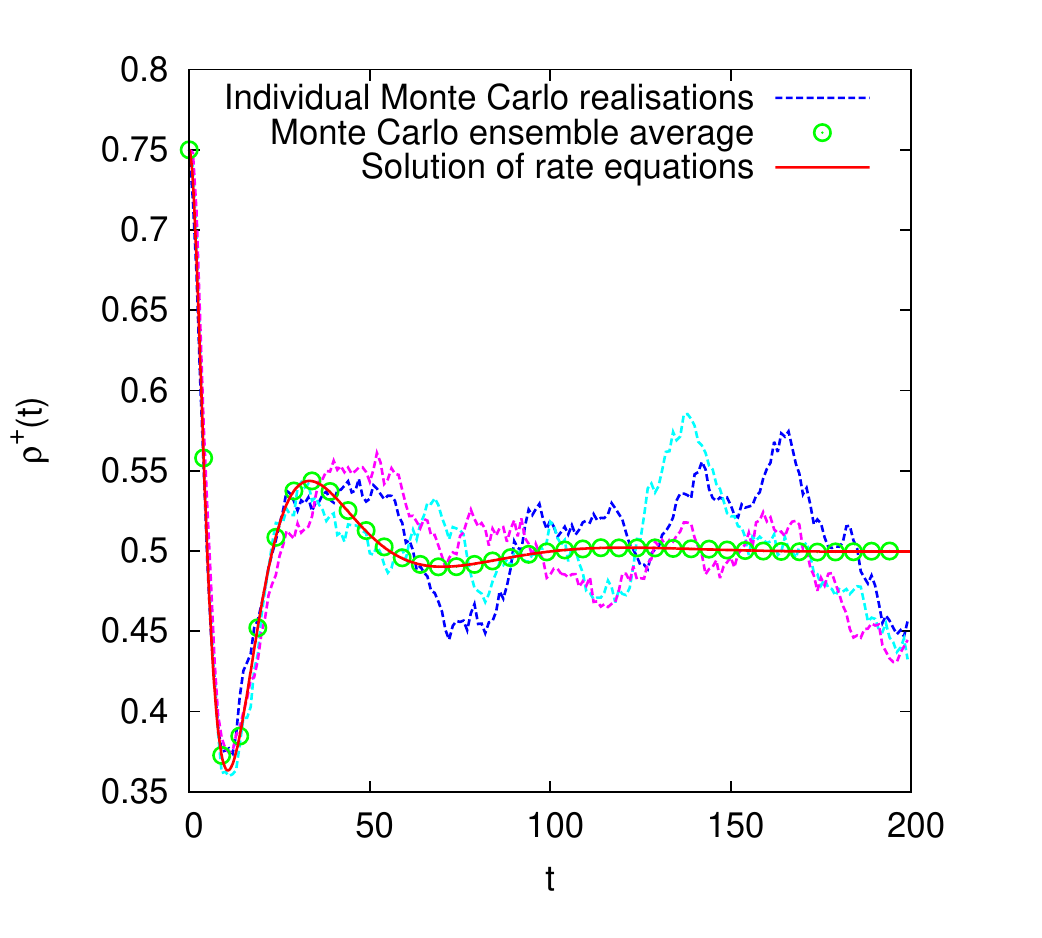}
\caption{Comparison between the evolution of $\rho_+(t)$ obtained from
  the numerical solution of Eqs.~(\ref{eq-MF}) (solid line) with that
  obtained from an ensemble average of $1000$ Monte Carlo realizations
  (circles) of the stochastic process described in
  Sec.~\ref{sec-model} with $N=6400$ agents. The value of $p$ is $1.0$
  and the initial condition is $\rho_+(0) = 0.75$.  Three different
  individual realizations of the stochastic dynamics are also shown in
  dashed lines.}
\label{fig-validateMF}
\end{center}
\end{figure}

\subsection{Rate equations}
\label{MF-eqns}

Let us now write down the mean-field (MF) rate equations for the stochastic
dynamics described in Sec.~\ref{sec-model}.  Adopting some of the
notation from \cite{ben-naim_structure_2006}, the time evolution of
the fraction of agents with opinion $+$ and fitness $k$, $f_k^+$, is
given by
\begin{eqnarray}
\nonumber \dd{{f}^+_k}{t} &=& p\,\left(f_{k\!-\!1}^+ F_{k\!-\!1}^- -
f_k^+F_k^- \right) + (1\!-\!p)\,\left(f_{k\!-\!1}^+ G_{k\!-\!1}^- -
f_k^+G_k^- \right)\\ \nonumber  &+& p\,\left(f_{k}^- G_{k}^+ -
f_k^+G_k^- \right) + (1\!-\!p)\,\left(f_{k}^- F_{k}^+ - f_k^+F_k^-
\right)\\
\label{eq-MF} &+& \frac{1}{2} \left(f_{k\!-\!1}^+ f_{k\!-\!1}^- -
f_k^+f_k^- \right), \\ 
\nonumber \dd{{f}^-_k}{t} &=& \left[ + \leftrightarrow -\right],
\end{eqnarray}
where
\begin{equation}
F_k^\pm = \sum_{i=0}^{k-1} f_i^\pm \mbox{\ \ \ and \ \ \ } G_k^\pm =
\sum_{i=k+1}^{\infty} f_i^\pm.
\label{eq-FG}
\end{equation}
The corresponding equation for $f_k^-$ is obtained by switching the
$+$ and  $-$ labels in Eq.~(\ref{eq-MF}). This occurs frequently in
the analysis which follows.  For the sake of conciseness,  we shall
not explicitly write out the symmetric partner  of each equation
unless it is necessary. All terms appear in pair of positive and
negative terms describing the gain and loss of agents having opinion
$+$ and $k$--value, $k$. The first pair of terms accounts for
interactions of $+$ agents with $-$ agents  having a lower $k$--value,
with the result that the original agent remains $+$. The second pair
of terms accounts for interactions of $+$ agents with  $-$ agents
having a higher $k$--value, with the result that the original agent
remains $+$. The third pair of terms accounts for the interactions of
$-$ agents with $+$ agents having a higher $k$--value, with the result
that the original agent switches opinion to $+$. The fourth pair of
terms   accounts for the interactions of $-$ agents with $+$ agents
having a lower $k$--value,  with the result that the original agent
switches opinion to $+$. The final pair of terms accounts for the
cases when two interacting agents have equal $k$--values.

The fraction of the total population in each group is
\begin{equation}
\label{eq-rho}
\rho_\pm (t) = \sum_{k=0}^\infty f_k^\pm.
\end{equation}
The mean $k$--value of each group is
\begin{equation}
\label{eq-mu}
\mu_\pm (t) = \sum_{k=0}^\infty k\,f_k^\pm.
\end{equation}
Since $f^\pm_k$ are proportions, we must have $\rho_+(t) +
\rho_-(t)=1$. We shall  denote the mean $k$--value across the entire
population by $\mu(t) = \mu_+(t)+\mu_-(t)$.  By summing
Eqs.~(\ref{eq-MF}) over $k$, we obtain the following equations for the
populations of each group:
\begin{eqnarray}
\nonumber \dd{\rho_+}{t} &=& (2 p -1)\,\sum_{k=0}^\infty
\left(f_k^+F_k^- - f_k^- F_k^+ \right), \\
\label{eq-rhoDot}\dd{\rho_-}{t} &=& \left[ + \leftrightarrow -\right].
\end{eqnarray}
It is clear from these formulae that the sum of the populations is
conserved.  Furthermore, when $p=\frac{1}{2}$ we see that the two
populations are conserved individually. This is to be expected since
for $p=\frac{1}{2}$ the dynamics of the $k$--counter is entirely
decoupled from the dynamics of exchange between the two opinion
groups.  Then, the competition between the groups is described by the
simple voter  model, for which we know that the individual populations
are conserved on average. By multiplying Eqs.~(\ref{eq-MF}) by $k$ and
summing over $k$, we obtain the following  less elegant equations for
the mean $k$--value of each group:
\begin{eqnarray}
\nonumber \dd{\mu_+}{t} &=& (1-p)\,\rho_+\rho_- + p\,\left(\rho_+\mu_-
- \rho_-\mu_+ \right) \\ \nonumber & &+ (2 p\!-\!1)\sum_{k=0}^\infty
k\,\left(f_k^+F_k^- - f_k^- F_k^+ \right)\\ \nonumber & &+ (2
p\!-\!1)\sum_{k=0}^\infty f_k^+F_k^-  + \frac{1}{2} (2
p\!-\!1)\sum_{k=0}^\infty f_k^+f_k^-, \\
\dd{\mu_-}{t} &=& \left[ + \leftrightarrow -\right].
\label{eq-muDot}
\end{eqnarray}
The average $k$--value of the whole population satisfies
\begin{eqnarray}
\dd{\mu}{t} &=& (2 p -1) \sum_{k=0}^\infty \left(  f_k^+F_k^-
+\frac{1}{2} f_k^+ f_k^- + f_k^-F_k^+\right) \nonumber \\ & &+ 2
(1-p)\,\rho_+\rho_- \nonumber \\ &=& \rho_+\,\rho_-,
\label{dudt}
\end{eqnarray}
where some careful algebra is required to establish simplify the sums.
Equation~(\ref{dudt}) shows that the mean of the total fitness grows
at a rate $\rho_+ \rho_-$, which is time dependent.  This equation can
also be derived by considering the mean change of $\mu$, $\Delta \mu$,
in a single time step $\Delta t=2/N$ of the dynamics.  An update
occurs only when a $+$ and a $-$ agents are chosen, which happens
with probability $2 \rho_+ \rho_-$.  Then, one of the agents increases
its fitness by one, changing $\mu$ by $1/N$ ($\Delta \mu=1/N$).  Then
we can write 
\begin{eqnarray}
\frac{d\mu}{dt} \simeq \frac{\Delta \mu}{\Delta t} = \frac{2 \rho_+
  \rho_- (\frac{1}{N})}{2/N}= \rho_+ \rho_-,
\end{eqnarray}
as in Eq.~(\ref{dudt}).  It is difficult to tell a-priori how the
total $k$--value in the system will behave since it depends on the
fraction of the population in the two groups. We note, however, that
when $p=1/2$, $\mu(t)$ grows linearly since $\rho_\pm$ are
constant and given by their initial values.  We can also check that
for $p=1/2$ and initial densities $\rho_+(0)=3/4$ and
$\rho_-(0)=1/4$, Eqs.~(\ref{eq-muDot}) are reduced to  
\begin{eqnarray}
\nonumber \dd{\mu_+}{t} &=& \frac{3}{32} + \frac{1}{8} (3 \mu_--\mu_+),
\\ \nonumber 
\dd{\mu_-}{t} &=& \frac{3}{32} - \frac{1}{8} (3 \mu_--\mu_+),
\label{eq-muDot-1}
\end{eqnarray}
whose solutions with initial mean $k$--values $\mu_+(0)=\mu_-(0)=0$ are 
\begin{eqnarray}
\mu_+(t) &=& \frac{3}{32} \left( \frac{3}{2} t + e^{-t/2} - 1 \right),
\nonumber \\ 
\mu_-(t) &=& \frac{3}{32} \left( \frac{1}{2} t - e^{-t/2} + 1 \right). 
\label{eq-muDot-2}
\end{eqnarray}
Equations~(\ref{eq-muDot-2}) show that the mean $k$--values increase
linearly at large times, as $\mu_+(t) \simeq 9 t/64$ and $\mu_-(t) \simeq 3
t/64$, as we can also see in the inset of Fig.~\ref{fig-MF_p=0.50}(A).

\subsection{Numerical solutions of the rate equations}

On a complete graph, Eqs.~(\ref{eq-MF}) exactly describe the ensemble
averaged behaviour of the system.  This is illustrated in
Fig.~\ref{fig-validateMF} which compares the evolution of
$\rho_+(t)$ (solid line) obtained from a numerical solution of
Eqs.~(\ref{eq-MF}) with the evolution of the ensemble average
of $\rho_+(t)$ (circles) obtained from $1000$ realisations of the
stochastic model described in Sec.~\ref{sec-model}.  Results
correspond to $p=1.0$ and initial condition $\rho_+(0)=0.75$.  Two features of
the dynamics are striking. Firstly, we notice that starting from an
asymmetric initial state that favors the $+$ opinion group ($\rho_+(0)=0.75$ and
$\rho_-(0)=0.25$), the dynamics drives the system towards a
coexistence state composed by even fractions of agents with $+$ and $-$
opinions ($\rho_+(t)=\rho_-(t) = 1/2$).  Secondly, we notice that the
approach to the coexistence state is not monotonic: the dynamics have a damped
oscillatory character. This suggests that the state in which both
populations are equal is a fixed point. Figure~\ref{fig-validateMF}
also shows three independent realisations (dashed lines), as compared
to the ensamble average.  This illustrates the importance of
fluctuations which are ultimately responsible for the system reaching
consensus as found in Sec.~\ref{sec-consensusTimes}, despite the fact
that the dynamics drives the system towards coexistence. 

\begin{figure}
\begin{center}
${\mathbf p=\frac{1}{4}}$ \subfigure{
    \includegraphics[width=0.9\columnwidth]{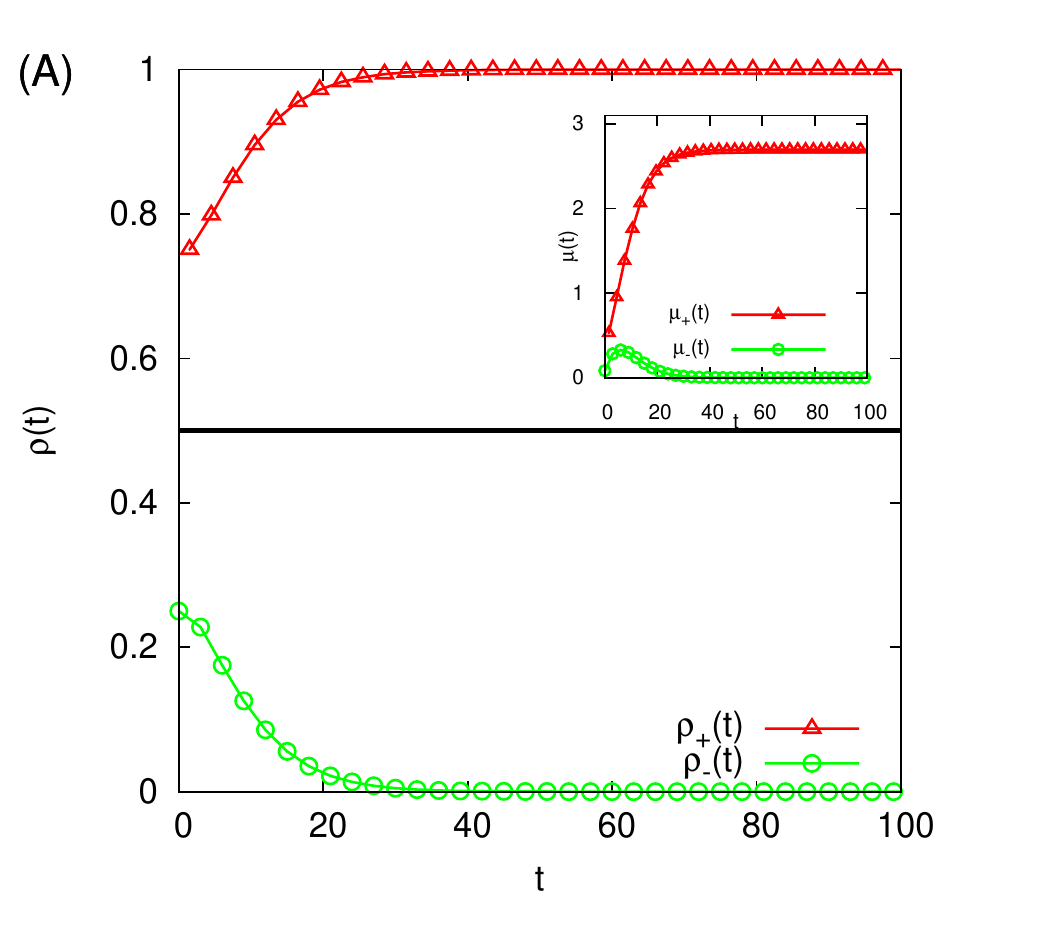} }
  \subfigure{
    \includegraphics[width=0.9\columnwidth]{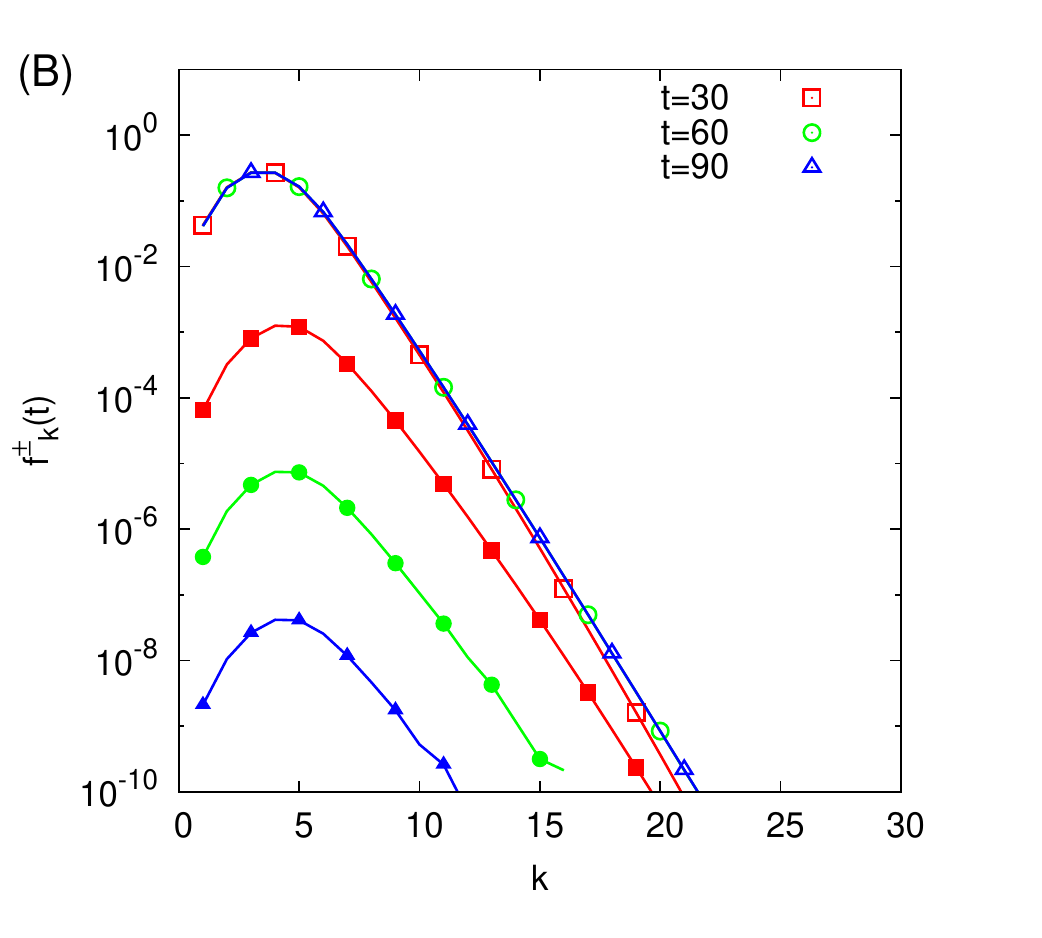} }
\caption{\label{fig-MF_p=0.25}(Color online) MF dynamics from
  Eqs.~(\ref{eq-MF}) for $p=1/4$.  (A)  Time evolution of the fraction
  of agents in the two opinion groups   $\rho_\pm(t)$.  Inset:
  Evolution of the mean fitness values $\mu_\pm(t)$.  (B) Snapshots of
  the $k$--value distributions $f_k^+$ (empty symbols) and $f_k^-$
  (filled symbols) at different times.}
\end{center}
\end{figure}

\begin{figure}
\begin{center}
${\mathbf p=\frac{1}{2}}$ \subfigure{
    \includegraphics[width=0.9\columnwidth]{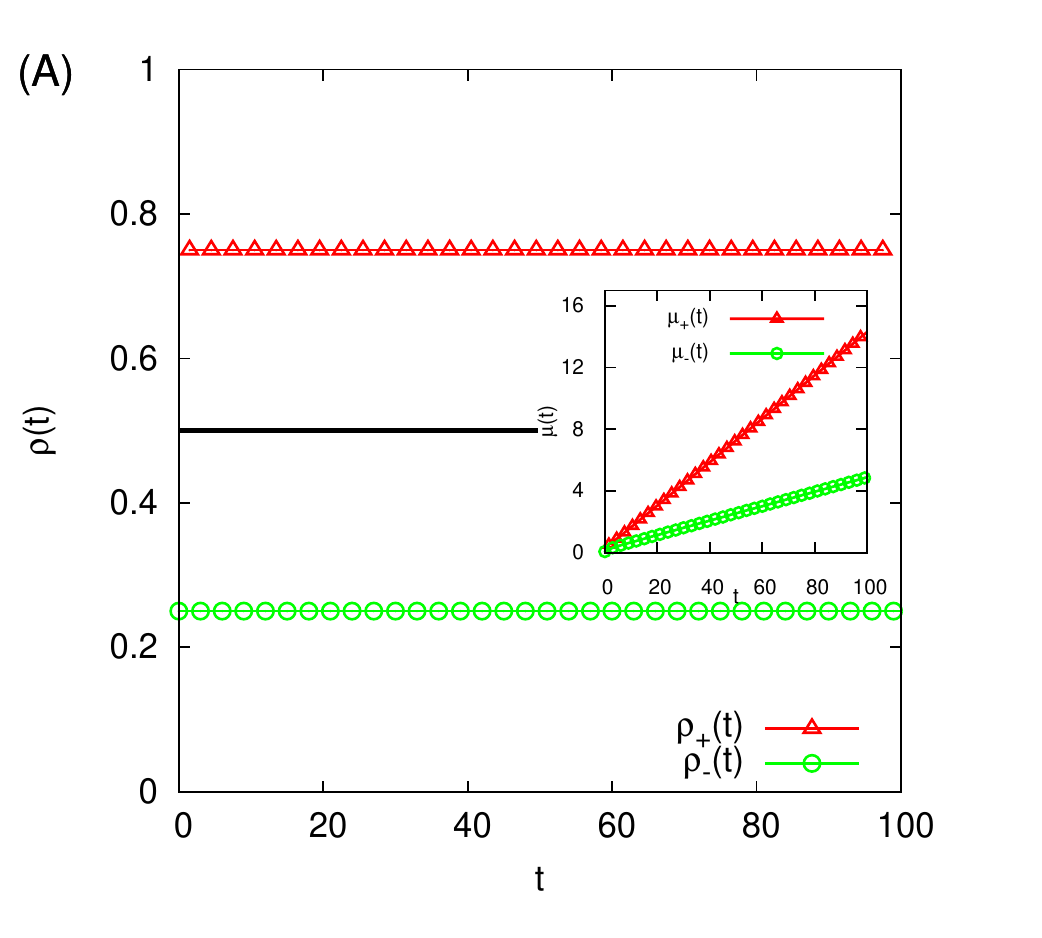} }
  \subfigure{
    \includegraphics[width=0.9\columnwidth]{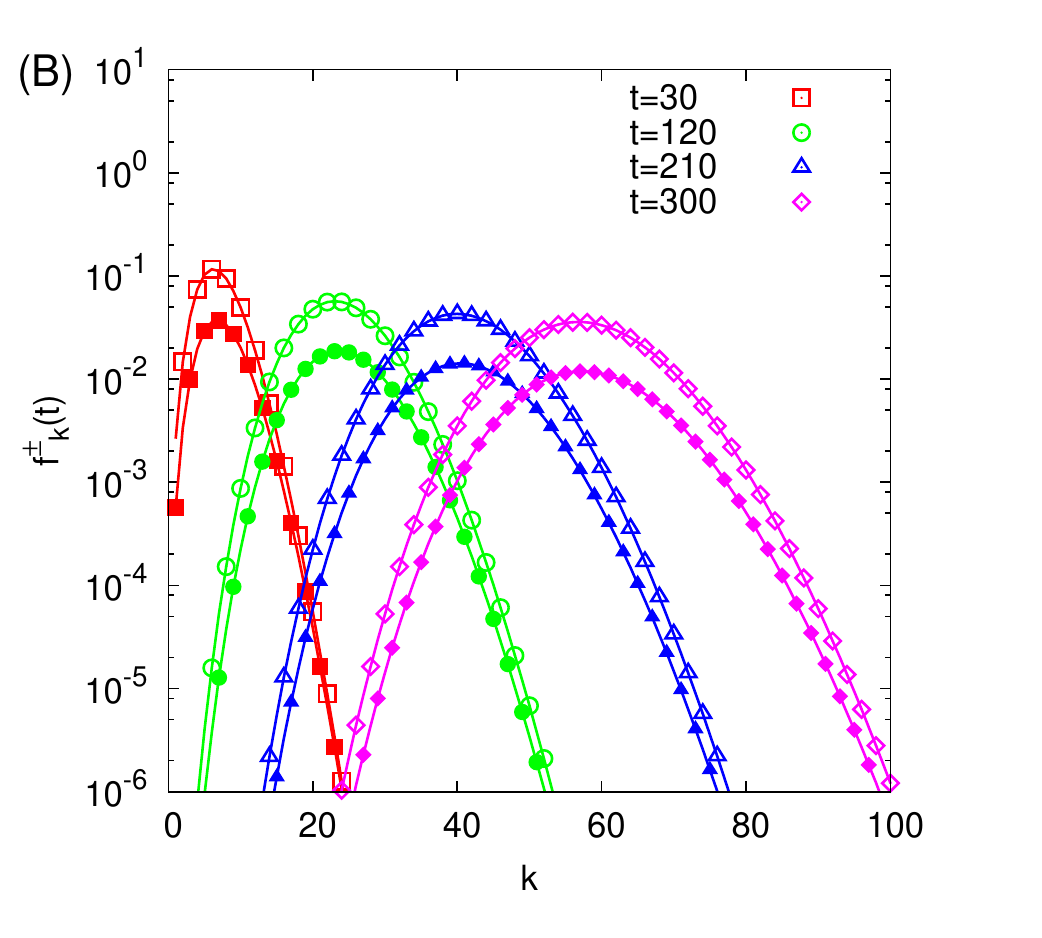}
  }
\caption{\label{fig-MF_p=0.50}(Color online) MF dynamics from
  Eqs.~(\ref{eq-MF}) for $p=1/2$.  (A)  Time evolution of the fraction
  of agents in the two opinion groups   $\rho_\pm(t)$.  Inset:
  Evolution of the mean fitness values $\mu_\pm(t)$.  (B) Snapshots of
  the $k$--value distributions $f_k^+$ (empty symbols) and $f_k^-$
  (filled symbols) at different times.}
\end{center}
\end{figure}

\begin{figure}
\begin{center}
${\mathbf p=\frac{3}{4}}$ \subfigure{
    \includegraphics[width=0.9\columnwidth]{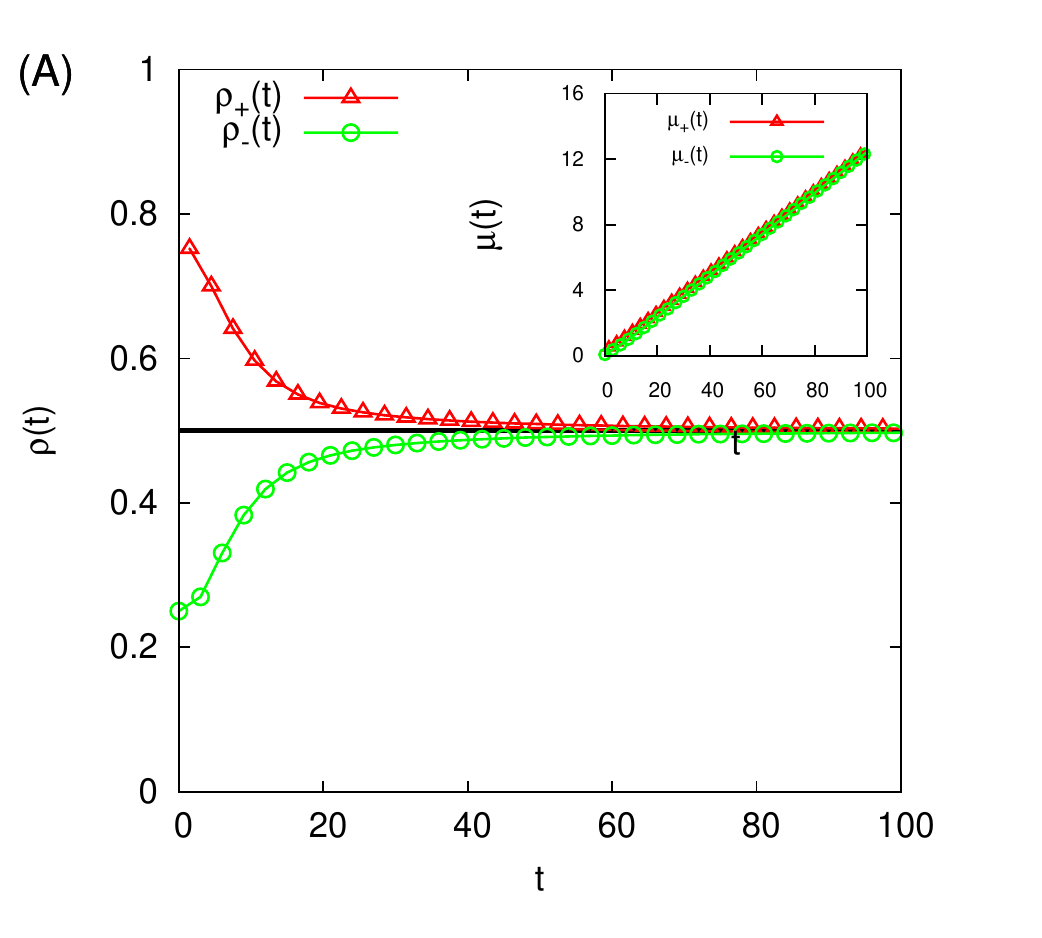}
  } \subfigure{
    \includegraphics[width=0.9\columnwidth]{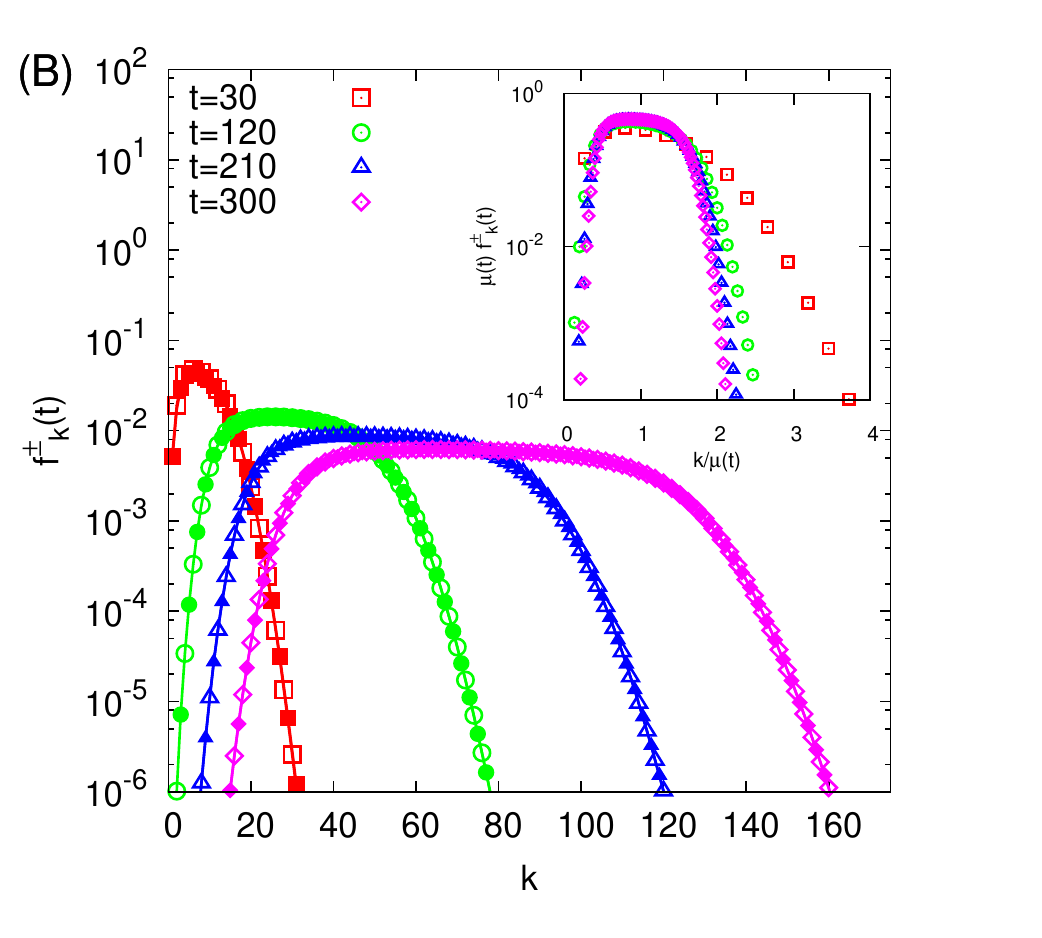}
  }
\caption{\label{fig-MF_p=0.75}(Color online) MF dynamics from
  Eqs.~(\ref{eq-MF}) for $p=3/4$.  (A)  Time evolution of the fraction
  of agents in the two opinion groups   $\rho_\pm(t)$.  Inset:
  Evolution of the mean fitness values $\mu_\pm(t)$.  (B) Snapshots of
  the $k$--value distributions $f_k^+$ (empty symbols) and $f_k^-$
  (filled symbols) at different times.  The inset shows the collapse of the
  data obtained from the self-similar scaling, Eq.~(\ref{eq-selfSimilarity}).}
\end{center}
\end{figure}

\begin{figure}
\begin{center}
${\mathbf p=1}$ \subfigure{
    \includegraphics[width=0.9\columnwidth]{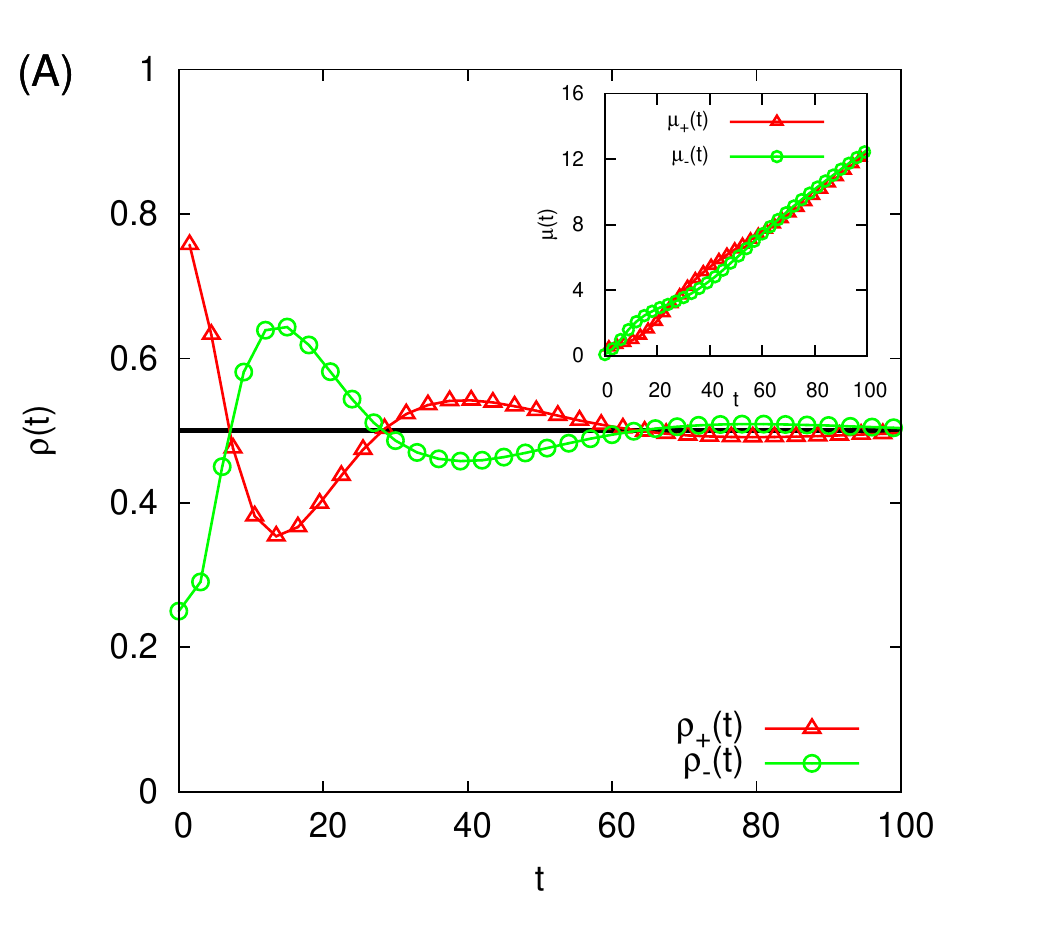} }
  \subfigure{
    \includegraphics[width=0.9\columnwidth]{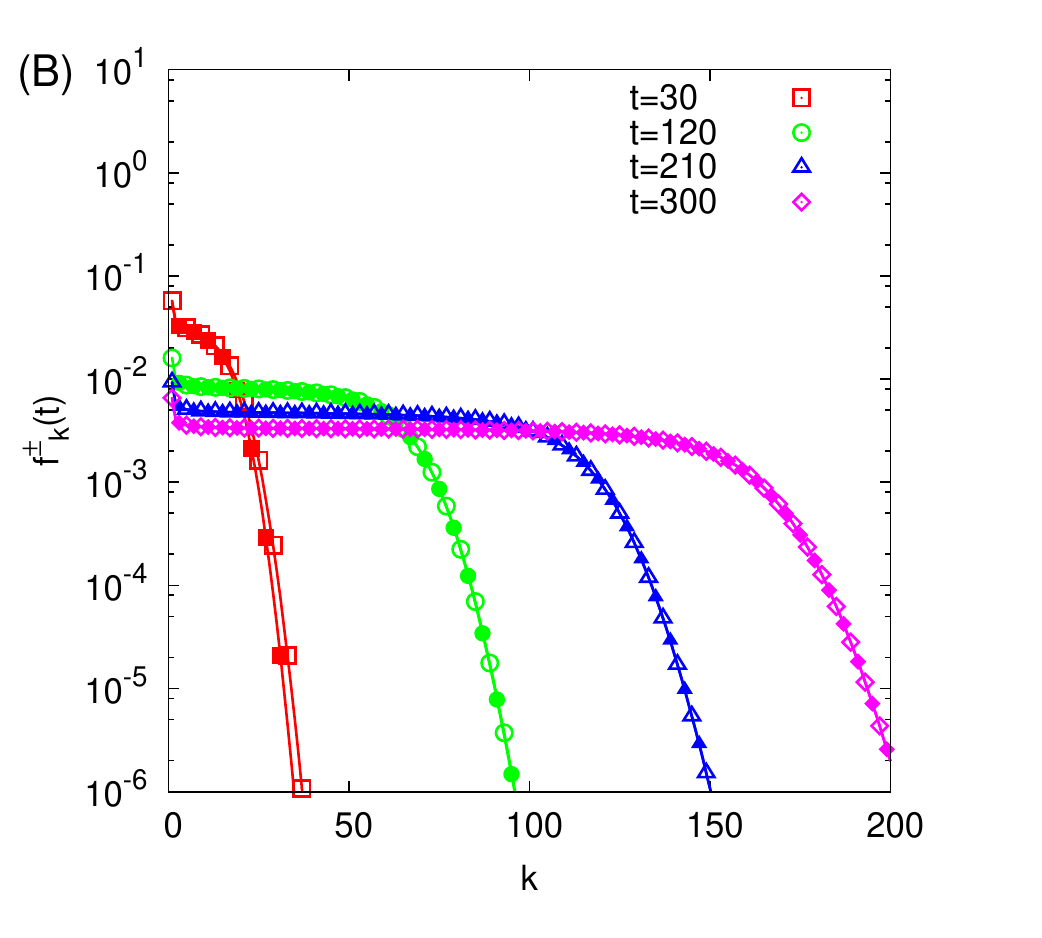}
  }
\caption{\label{fig-MF_p=1.00}(Color online) MF dynamics from
  Eqs.~(\ref{eq-MF}) for $p=1$.  (A)  Time evolution of the fraction
  of agents in the two opinion groups   $\rho_\pm(t)$.  Inset:
  Evolution of the mean fitness values $\mu_\pm(t)$.  (B) Snapshots of
  the $k$--value distributions $f_k^+$ (empty symbols) and $f_k^-$
  (filled symbols) at different times.}
\end{center}
\end{figure}

In Figs.~\ref{fig-MF_p=0.25}--\ref{fig-MF_p=1.00} we explore  the
behaviour of the system by means of Eqs.~\ref{eq-MF} in the entire
range of $p$ values: $p=1/4$ (Fig.~\ref{fig-MF_p=0.25}), $p=1/2$
(Fig.~\ref{fig-MF_p=0.50}),  $p=3/4$ (Fig.~\ref{fig-MF_p=0.75}) and
$p=1$ (Fig.~\ref{fig-MF_p=1.00}).
Figures~\ref{fig-MF_p=0.25}(A)--\ref{fig-MF_p=1.00}(A) show the time
evolution of $\rho_\pm(t)$, while their insets show the dynamics of
$\mu_\pm(t)$.  Figures~\ref{fig-MF_p=0.25}(B)--\ref{fig-MF_p=1.00}(B)
show snapshots at various times of the corresponding $k$--value
distributions, $f^\pm_k(t)$.  In Fig.~\ref{fig-MF_p=0.25} we see that
for $p=1/4$ the system quickly reaches consensus in the opinion of the
initial majority, as $\rho_+$ approaches exponentially fast to $1.0$.
This is in agreement with the fast consensus observed in 
Fig.~\ref{fig-consensusTimes}(B) for $p<1/2$.  Thus, the dynamics for $p<1/2$
favors the opinion state of the majority, creating a positive 
feedback in which the largest opinion group permanently increases while the
smallest group shrinks and eventually disappear.  This can also be
seen in Fig.~\ref{fig-MF_p=0.25}(B), where $f^-_k(t)$ vanishes for
long times (filled symbols).  Figure~\ref{fig-MF_p=0.50}(A) shows that
the fractions $\rho_\pm$ are conserved for $p=1/2$, and that the
mean $k$--values $\mu_\pm$ grow linearly with time, as discussed in
Sec.~\ref{MF-eqns}.  For both $p=3/4$ and $p=1$ the system quickly
relaxes to the coexistence state $\rho_+=\rho_-=1/2$, but in different
ways.  Whereas for $p=3/4$ densities $\rho_\pm$ decay monotonically
towards the value $1/2$, for $p=1$ the approach to coexistence
exhibits damped oscillations.  Finite-size fluctuations --not captured
by the MF equations-- eventually drive the system from the
coexistence state to consensus, which leads to the very long consensus
times measured for $p>1/2$ [see Fig.~\ref{fig-consensusTimes}(A)].

\subsection{$p>\frac{1}{2}$ : Self-similar solution for the coexistence state}

The above  argument requires that for $p>1/2$ the large time solution
of Eqs.~(\ref{eq-MF}) converge to a coexistence state in which both 
populations have the same fraction of agents ($\rho_+=\rho_-$) and 
equivalent $k$--value distributions, as we can see in
Figs.~\ref{fig-MF_p=0.75} and \ref{fig-MF_p=1.00}.  In this section
we demonstrate the existence of a self-similar solution of
Eqs.~(\ref{eq-MF}) which achieve this. Substituting $f_k^+(t) =
f_k^-(t) = f_k(t)$ in these equations and simplifying gives:
\begin{align}
\nonumber \frac{d}{dt}{f}_{k}  &= p \left( f_{k - 1} F_{k - 1} - f_{k}
F_{k} \right) + (1- p) \left( f_{k - 1} G_{k - 1} - f_{k} G_{k}
\right)\\ & +  \frac{1}{2} \left( f_{k - 1}^2 - f_{k}^2 \right).
\label{eq:dyneqnsingle}
\end{align}
Following Ben--Naim et al. \cite{ben-naim_structure_2006} we rewrite
this equation using $f_k = F_{k+1} - F_{k}$ and $G_k = F_{\infty} -
F_{k+1} = 1/2 - F_{k+1}$ (where the maximum group size is $\rho_+ =
F^+_{\infty} = 1/2$ or $\rho_- = F^-_{\infty} = 1/2$). This gives a
closed equation for the  cumulative distribution:
\begin{align}
\nonumber \frac{d}{dt}{F}_{k}  &= p F_{k - 1} ( F_{k - 1} - F_{k} ) -
\frac{1}{2} \left( F_{k} - F_{k-1} \right)^2\\ & + (1- p) ( 1/2 - F_k
) ( F_{k - 1} - F_{k} ).
\label{eq:dyneqnCumulsingle}
\end{align}
The boundary conditions are $F_{0} = 0$ and $F_{\infty}=1/2$. Taking
the continuum limit of Eq. (\ref{eq:dyneqnCumulsingle}) and keeping
only the leading order term, we get
\begin{equation}
\frac{\partial F}{\partial t} =  \left[ \frac{1}{2}(p-1) + (1-2p)F
  \right] \frac{\partial F}{\partial k}.
\label{eq:dyneqnCumulsingle2}
\end{equation}
We have observed already that $\mu_\pm(t)$ grows linearly for large
times when $p>1/2$, thus the mean value of the distributions 
\begin{eqnarray*}
\langle k \rangle^{\pm} = \frac{1}{\rho_\pm} \sum_{k=0}^\infty k\,f_k^\pm =
\frac{\mu_\pm}{\rho_\pm}=2 \mu_\pm=\mu
\end{eqnarray*}
also increase linearly with time.  This suggests that we look for solutions of
Eq.~(\ref{eq:dyneqnCumulsingle2}) in which the cumulative
distribution, $F(k,t)$, has the scaling form:
\begin{equation}
F(k,t) \simeq \Phi (\xi)\hspace{0.5cm}\mbox{with $\xi \equiv
  \frac{k}{\mu(t)}=\frac{k}{\alpha t}$},
\end{equation}
where $\alpha \simeq 0.24$ is twice the slope of the $\mu_\pm$ vs
$t$ curves for $t>60$ in the inset of Figs.~\ref{fig-MF_p=0.75} and
\ref{fig-MF_p=1.00}.  This is equivalent to the following self-similar
form for the $k$--value distribution itself
\begin{equation}
\label{eq-selfSimilarity}
f^\pm_k(t) \simeq \mu(t)^{-1}\, \phi(\xi)\hspace{0.5cm}\mbox{where
  $\varphi \equiv \frac{d\Phi}{d \xi}$.}
\end{equation}
In the scaling variables, Eq.~(\ref{eq:dyneqnCumulsingle2}) takes the
form
\begin{equation}
\left[ \frac{1}{2}(p-1)+\alpha \, \xi + (1-2p)\Phi(\xi)\right]\frac{d
  \Phi}{d\xi} = 0,
\label{eq:scaled1}
\end{equation}
Solving Eq.~(\ref{eq:scaled1}) gives 
\begin{displaymath}
\Phi(\xi)=\mathrm{constant} \hspace{0.5cm}\mbox{or}\hspace{0.5cm}
\Phi(\xi) = \frac{\frac{1}{2}(p-1)}{2p-1} + \frac{\alpha \, \xi}{2p-1}.
\end{displaymath}
Following \cite{ben-naim_structure_2006}, we use the boundary
conditions $\Phi(0)=0$ and $\Phi(\infty)=1/2$, the monotonicity of
$F_k$, $d \Phi/d\xi \ge 0$, and the bounds $0 \le \Phi \le 1$, to
assemble a sensible piece-wise smooth solution: 
\begin{equation} 
\label{eq-Phi}
\Phi(\xi) = 
\begin{cases} 
0 & \mbox{if   $0 < \xi < \xi_{-}$} \\  \frac{\frac{1}{2}(p-1)}{2p-1} +
\frac{\alpha \, \xi}{2p-1} & \mbox{if   $\xi_{-} < \xi < \xi_{+}$}
\\  1/2 & \mbox{if $\xi_{+} < \xi$,}
\end{cases}
\end{equation}
where $\xi_{-} = (1-p)/2 \alpha$ and $\xi_{+} = p/2 \alpha$. We remark
that if we keep the next order (diffusive) terms in the derivation of
Eq.~(\ref{eq:dyneqnCumulsingle2}), then the sharp corners in this
solution would be smoothed out. Differentiating this solution with
respect to $\xi$ gives the corresponding scaling function for the
$k$--value distribution itself [see Eq.~(\ref{eq-selfSimilarity})]
\begin{equation}
\label{eq-phi}
\mu(t) f_k^{\pm}(t) = \phi(\xi) = 
\begin{cases} 
0 & \mbox{if   $0 < \xi < \xi_{-}$} \\  \frac{\alpha}{2p-1} & \mbox{if
  $\xi_{-} < \xi < \xi_{+}$} \\  0 & \mbox{if   $\xi_{+} < \xi$,}
\end{cases}
\end{equation}
where $\xi_\pm$ are the same as in Eq.~(\ref{eq-Phi}) above. Note that
as $p\to 1/2$, this solution tends to a $\delta$-function. This case
will be considered in the next section. The inset of
Figs.~\ref{fig-MF_p=0.75}(B) and \ref{fig-MF_p=1.00}(B) shows the data
collapse of some snapshots of the full $k$--value distributions
obtained from numerics (symbols) onto the curve given by
Eq.~(\ref{eq-phi}) (solid line).  

\subsection{$p=\frac{1}{2}$: Dynamics of $k$--value distribution}

For completeness, let us look at what happens on the boundary when
$p=1/2$. Many of the terms in Eqs.~(\ref{eq-MF}) are then absent due
to the fact that $F_k^\pm(t) + f_k^\pm(t) + G_k^\pm(t) =
\rho_\pm(t)$. Further simplification follows from
Eq.~(\ref{eq-rhoDot}) which tells us that $\rho_\pm(t)$ are constant
when $p=1/2$. This reflects the fact that the evolution of the
distribution of $k$--values decouples from the opinion dynamics which
are equivalent to the standard voter model dynamics in which the average
magnetization is conserved. Eqs.~(\ref{eq-MF}) become
\begin{eqnarray*}
\pd{f^+_k}{t} &=&\frac{1}{2}\rho_-\left(f_{k-1}^+ - f_k^+\right) +
\rho_+f_k^- - \rho_-f_k^+\\ \pd{f^-_k}{t} &=&\left[ + \leftrightarrow
  -\right].
\end{eqnarray*}
Taking the continuum limit we get
\begin{eqnarray}
\nonumber \pd{f^+_k}{t} &=& -\frac{1}{2}\rho_-\pd{f_k^+}{k} + 
\frac{1}{4}\rho_-\pdd{f_k^+}{k} -\rho_-f_k^+ + \rho_+f_k^-\\
\label{eq-p0.5CtLimit}\pd{f^-_k}{t} &=&\left[ + \leftrightarrow -\right].
\end{eqnarray}
The initial conditions are $f_k^\pm=\rho_\pm\,\delta(k)$. We should
also impose the zero-flux boundary conditions, $\pd{f_k^\pm}{k}=0$ at
$k=0$ since the $k$--values cannot become
negative. Eqs.~(\ref{eq-p0.5CtLimit}) are a pair of coupled linear
equations which can be decoupled and solved in Fourier-space. The
symmetric case $\rho_+=\rho_-=1/2$ is particularly simple and
illustrative, in which both distributions are identical and evolve
according to the single advection-diffusion equation,
\begin{equation}
\frac{\partial f^\pm}{\partial t} = - \frac{1}{4} \frac{\partial
  f^\pm}{\partial k} + \frac{1}{8} \frac{\partial^2 f^\pm}{\partial k^2}.
\label{eq:P05eqnsAdvDiff}
\end{equation}
The solution of this equation on the whole of $\mathbb{R}$ is
\begin{displaymath}
 f^\pm(k,t) = {\sqrt\frac{2}{\pi t}}\,  \mathrm{e}^{- \frac{2\,(k -
     t/4)^2}{t}}.
\end{displaymath}
In order to satisfy the boundary condition at $k=0$ we can employ the
method of images to obtain
\begin{equation}
f^\pm(k,t) = \frac{1}{2}\sqrt\frac{2}{\pi t}\,\left(  \mathrm{e}^{-
  \frac{2\,(k - t/4)^2}{t}} + \mathrm{e}^{- \frac{2\,(k +
    t/4)^2}{t}} \right).
\end{equation}
We see that the $k$--value distribution is a Gaussian which propagates
to the larger values of $k$ with fixed speed $v=1/4$, and whose width
increases with time as $\sigma=\sqrt{t}/2$. This remains true,
although it is more difficult to show analytically if
$\rho_+ \neq \rho_-$ [see Fig.~\ref{fig-MF_p=0.50}(B)]. Note that by
including the higher order derivative in taking the continuous limit
in Eq.~(\ref{eq-p0.5CtLimit}), the singular behaviour of the scaling
solution, Eq.~(\ref{eq-phi}), is regularized.

\section{A reduced model of the dynamics}
\label{sec-reduced}

In order to better understand the dynamics described in
Sec.~\ref{sec-MF}, we  introduce a reduced model which captures most
of the essential features of  Eqs.~(\ref{eq-MF}) but is simple enough
to allow some insight to be obtained. If we think of the $f_k^\pm$ as
being analogous to probability distributions, then their
specification is equivalent to the specification of all their
moments. We have already seen, however, that even the first two
moments, $\rho^\pm$ and $\mu^\pm$, satisfy complicated equations,
Eqs.~(\ref{eq-rhoDot}) and (\ref{eq-muDot}) involving
cross-correlations between $f_k^+$ and $f_k^-$. In principle, one
could use the dynamical equations to write evolution equations for
these cross-correlations but such equations would involve triple
correlations and so on. Such an approach is unlikely to lead
anywhere. Instead, in the spirit of moment closures and  single point
closures in turbulence, we suggest to close the system at the level of
the first order (in $f_k^\pm$) quantities $\rho^\pm$ and
$\mu_\pm$. That is to say, we attempt to "approximate" the RHS of
Eqs.~(\ref{eq-rhoDot}) and (\ref{eq-muDot}) with functions of
$\rho_\pm$ and $\mu_\pm$ only. This would yield a simple three
dimensional dynamical system (not four dimensional because
$\rho_++\rho_-=1$) in place of the infinite hierarchy of equations in
(\ref{eq-MF}). Of course,  this cannot be done exactly and the trick
in obtaining useful closures is to come up with a reasonable proposal
for these functions should be. 

Inspired by the $k$--value distributions in
Fig.~\ref{fig-MF_p=0.75}(C) and aiming to build the simplest possible
model, we introduce the following model for the $k$--value
distributions:
\begin{eqnarray}
\label{eq-ftilde}
\widetilde{f}_k^\pm
= \frac{\rho_\pm^2}{2 \mu_\pm}\, \, \Theta(k) \, 
\Theta\left(\frac{2 \mu_\pm}{\rho_\pm}- k\right), 
\end{eqnarray}
where $\Theta(x)$ is the Heaviside theta function. We therefore treat
the distributions $f_k^\pm$ as being uniform on an interval
$\left[0,K\right]$.  The width, $K$, of this interval and the value of
the function on the interval are  chosen such that we have the
following properties:
\begin{eqnarray*}
\int_0^\infty \widetilde{f}_k^\pm\,dk &=& \rho_\pm\\
\int_0^\infty  k\,\widetilde{f}_k^\pm\,dk &=& \mu_\pm,
\end{eqnarray*}
where, in order to simplify things, we shall treat the $k$--value as a
continuous variable from this point on. We can now integrate
Eq.~(\ref{eq-ftilde}) to get a model for the cumulative distribution,
$F_k^\pm$:
\begin{equation} 
\label{eq-Ftilde} 
\widetilde{F}_k^\pm = 
\begin{cases} 
\frac{\rho_{\pm}^2}{2 \mu_{\pm}} k & \mbox{if   $0 \le
k \le \frac{2 \mu_{\pm}}{\rho_{\pm}}$}, \\  
\rho_{\pm} & \mbox{if $k \ge \frac{2 \mu_{\pm}}{\rho_{\pm}}$.}
\end{cases}
\end{equation}

We now substitute Eqs.~(\ref{eq-ftilde}) and (\ref{eq-Ftilde}) into
Eqs.~(\ref{eq-rhoDot}) and (\ref{eq-muDot}) and perform the
integrations on the RHS in order to obtain  expressions which depend
only on $\rho_\pm$ and $\mu_\pm$. This is a surprisingly tedious
process given the deceptive simplicity of Eq.~(\ref{eq-ftilde}). Using
{\em Mathematica} and massaging the output a little, we obtained the
following dynamical system
\begin{subequations}
\begin{alignat}{4}
\label{eq-reducedrp} \dd{\rho_+}{t} &=(2 p - 1)\,R^+_1,\\
\label{eq-reducedrm}
\dd{\rho_-}{t} &= \left[ + \leftrightarrow -\right],\\
\nonumber \dd{\mu_+}{t} &=  (1-p)\,\rho_+\rho_- +
p\,\left(\rho_+\mu_- - \rho_-\mu_+ \right)\\ 
\label{eq-reducedmp} &+(2 p\!-\!1) \left(R^+_2+R^+_3+R^+_4 \right),\\
\label{eq-reducedmm}\dd{\mu_-}{t} &= \left[ + \leftrightarrow -\right],
\end{alignat}
\label{eq-reduced}
\end{subequations}
where
\begin{subequations}
\begin{alignat}{4}
R^+_1 &=  \int_0^\infty \left(\widetilde{f}_k^+\widetilde{F}_k^-
- \widetilde{f}_k^- \widetilde{F}_k^+ \right)\,dk \\ &=
\left(\mu_+ \rho_- - \mu_- \rho_+ \right) R^+, \nonumber \\
R^+_2 &= \int_0^\infty k\,\left(\widetilde{f}_k^+\widetilde{F}_k^-
- \widetilde{f}_k^- \widetilde{F}_k^+ \right)\,dk \\ 
&= \frac{\left(\mu_+^2 \rho_-^2 - \mu_-^2 \rho_+^2 \right)
R^+}{\rho_+ \rho_-}, \nonumber \\
R^+_3 &= \int_0^\infty \widetilde{f}_k^+\widetilde{F}_k^-\,dk \\
&= \frac{1}{4 \mu_+ \mu_-} \Big\{ 2 \mu_+^2 \rho_-^2 \nonumber \\ 
&- \left(\mu_+ \rho_- - \mu_- \rho_+ \right) \left[ \left(\mu_+ \rho_-
- \mu_- \rho_+ \right) + |\mu_+ \rho_-
- \mu_- \rho_+| \right] \Big\}, \nonumber \\
\label{R4}
R^+_4 &= \frac{1}{2} \int_0^\infty \widetilde{f}_k^+\widetilde{f}_k^-\,dk = 
\frac{\rho_+ \rho_- R^+}{4},
\end{alignat}
\label{eq-Rs+}
\end{subequations}
with
\begin{equation}
R^+ = \frac{1}{2 \mu_+ \mu_-} 
\left[ \left(\mu_+ \rho_- + \mu_- \rho_+ \right) - 
|\mu_+ \rho_- - \mu_- \rho_+| \right].
\label{eq-R+}
\end{equation}
When performing the integrals we considered separately the two
cases $\mu_+ \rho_- > \mu_- \rho_+$ and $\mu_+ \rho_- < \mu_- \rho_+$,
which leaded to expressions defined by parts.  Then, we used the
absolute value function $|\bullet|$ to rewrite these expressions.  For
instance   
\begin{equation} 
R^+_4 = 
\begin{cases} 
R^+_a & \mbox{if $\mu_+ \rho_- > \mu_- \rho_+$}, \\  
R^+_b & \mbox{if $\mu_+ \rho_- < \mu_- \rho_+$,} 
\end{cases}
\end{equation}
with $R^+_a=\rho_+^2 \rho_-/4\mu_+$ and
$R^+_b=\rho_+ \rho_-^2/4\mu_-$, was rewritten as    
\begin{eqnarray*}
R^+_4= \frac{(R^+_a+R^+_b)}{2} + \frac{(R^+_a-R^+_b)}{2} 
\frac{|\mu_+ \rho_- - \mu_- \rho_+|}{(\mu_+ \rho_- - \rho_- \mu_+)},
\end{eqnarray*}
which is reduced to Eq.~(\ref{R4}) after some algebra.

This reduced model reproduces all the qualitative features of the full
MF equations, (\ref{eq-MF}). In particular, the absorbing
states in Eqs.~(\ref{eq-MF}) correspond to lines  of fixed points in
the reduced model. We refer to these as the $+$ and $-$ consensus fixed
points $P_+$ and $P_-$. They are parameterized by a single parameter, $\mu>0$:
\begin{eqnarray*}
P_+ &:& (\rho_+,\rho_-,\mu_+,\mu_-) = (1,0,\mu, 0)\\ P_- &:&
(\rho_+,\rho_-,\mu_+,\mu_-) = (0,1,0,\mu).
\end{eqnarray*}
That these points are zeroes of the right-hand side of
Eqs.~(\ref{eq-reduced})--(\ref{eq-R+}) for any value of $\mu$ can be verified by
direct substitution.   

\begin{figure*}
\begin{center}
\subfigure{
\includegraphics[width=0.9\columnwidth]{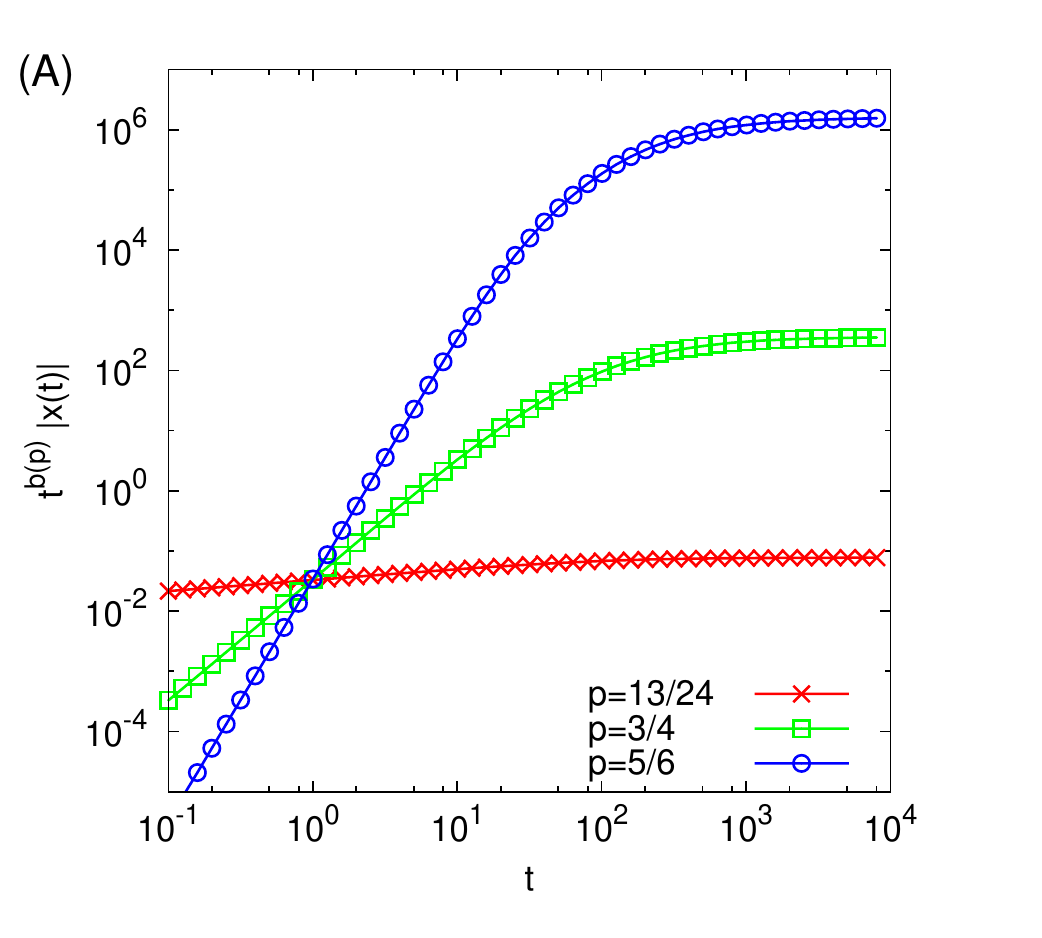}
\label{fig-closureVaryIC-x}}
\subfigure{
\includegraphics[width=0.9\columnwidth]{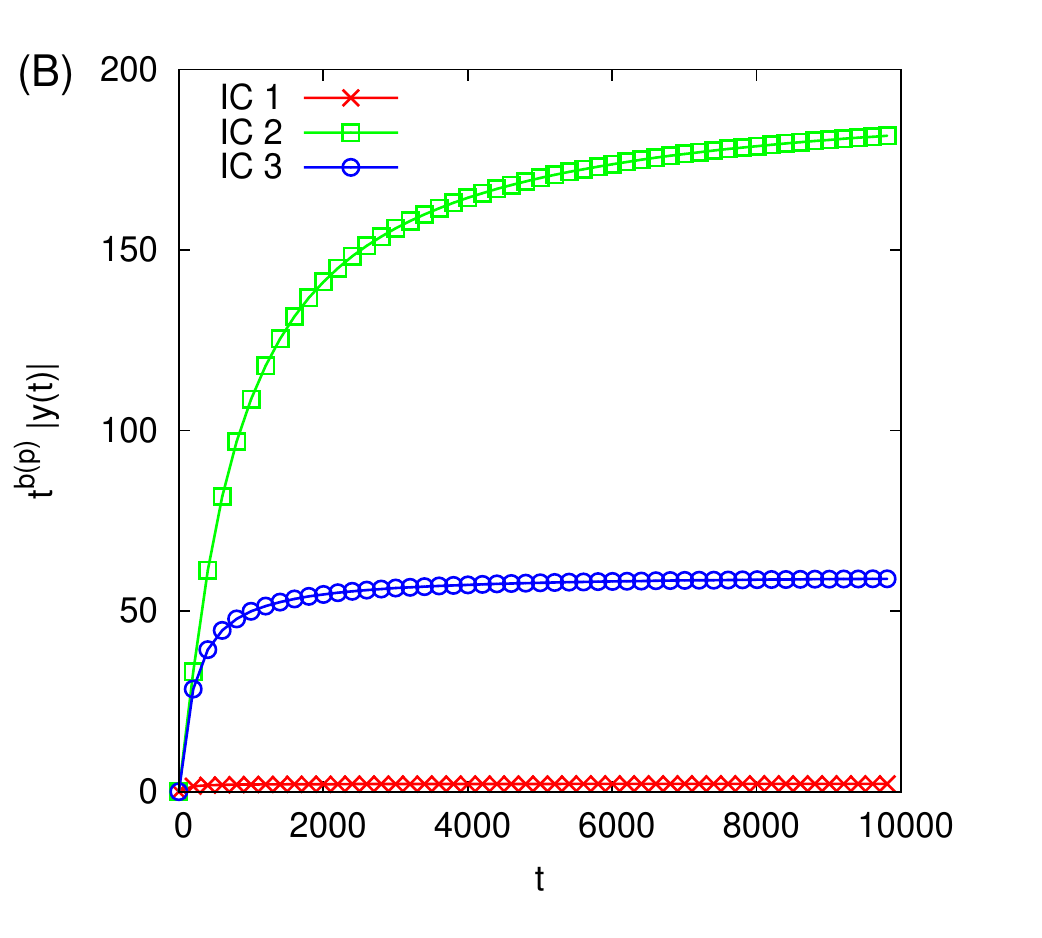}
\label{fig-closureVaryIC-y}}
\subfigure{
\includegraphics[width=0.9\columnwidth]{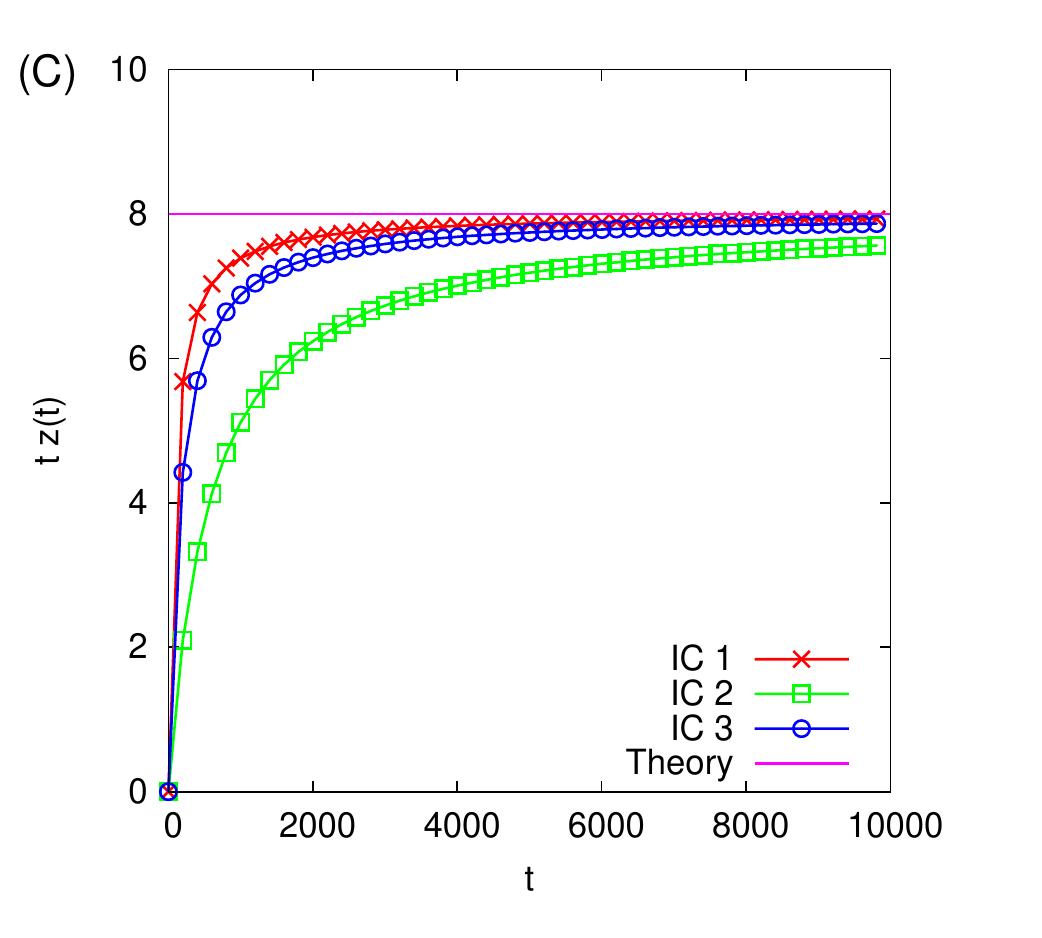}
\label{fig-closureVaryIC-z}}
\subfigure{
\includegraphics[width=0.9\columnwidth]{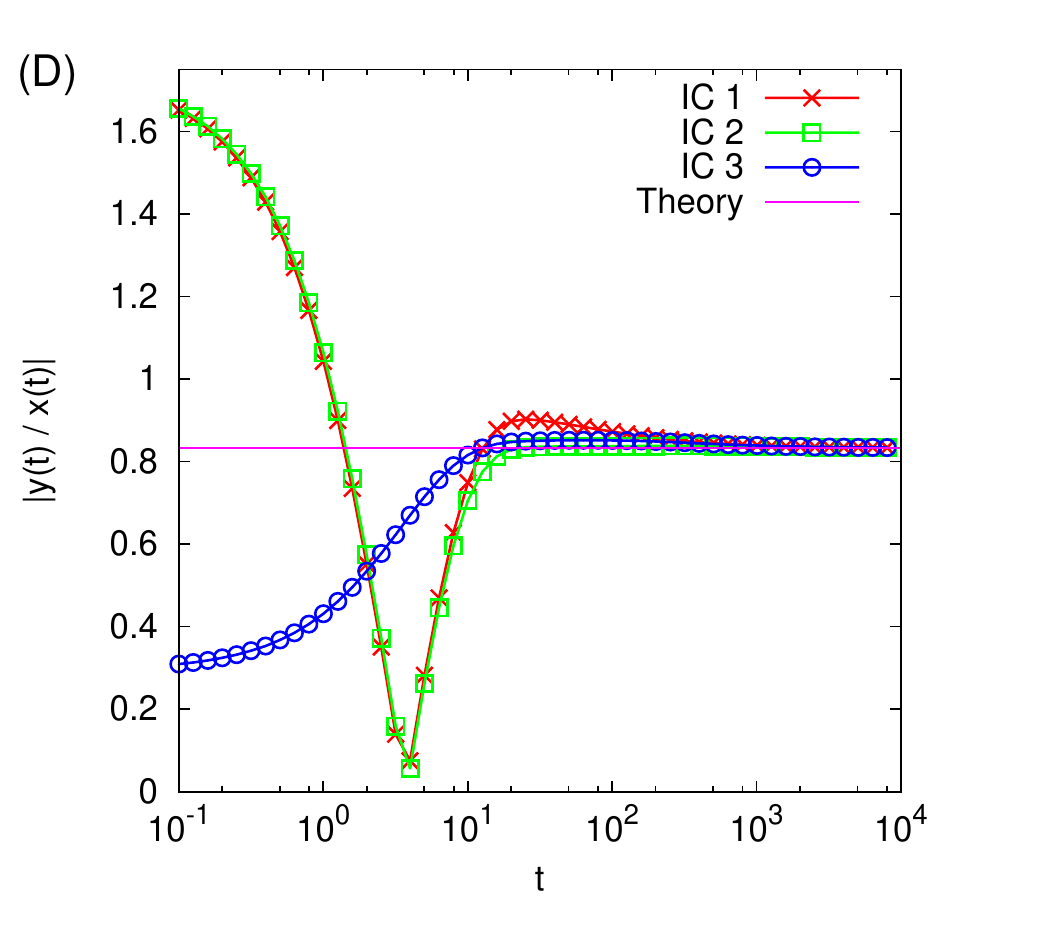}
\label{fig-closureVaryIC-ratio}}
\caption{\label{fig-closureVaryIC}(Color online) Numerical
investigation of the behaviour of the reduced model,
Eqs.~(\ref{eq-reducedXYZ}), near the coexistence fixed point, $P_0$,
for $p=\frac{7}{10}$. Panels (A), (B) and (C) show $\widetilde{X}(t)$, 
$Y(t)$ and $Z(t)$ respectively for three different generic initial
conditions. To make the agreement with theory clear, the data have
been compensated by the $t$-scalings predicted by
Eqs.~(\ref{eq-predX})-(\ref{eq-predZ})  with $b(p)$ given by
Eq.~(\ref{eq-bValue}). Panel (D) shows that the ratio
$Y(t)/\widetilde{X}(t)$ asymptotically approaches the theoretical
prediction $5/6$ (solid line) from Eq.~(\ref{eq-predRatio}).}
\end{center}
\end{figure*}

\begin{figure*}
\begin{center}
\subfigure{
\includegraphics[width=0.9\columnwidth]{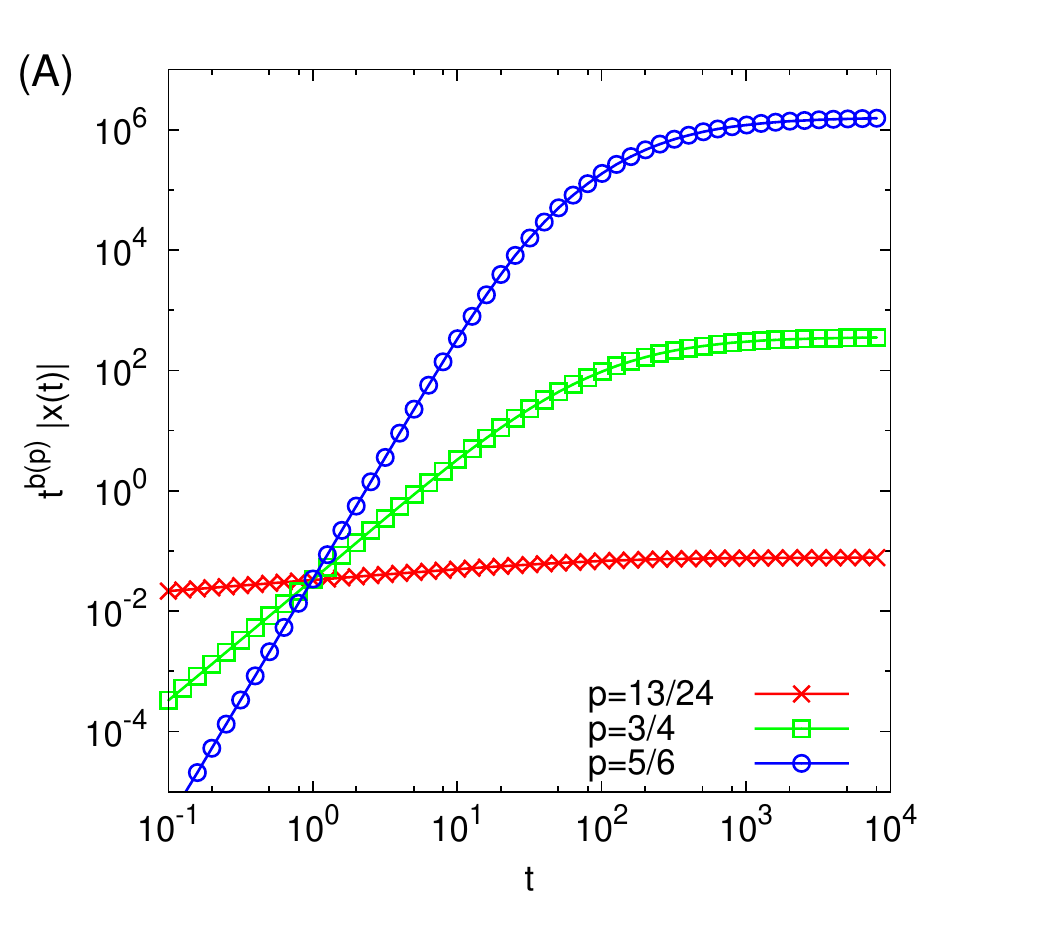}
\label{fig-closureVaryp-x}}
\subfigure{
\includegraphics[width=0.9\columnwidth]{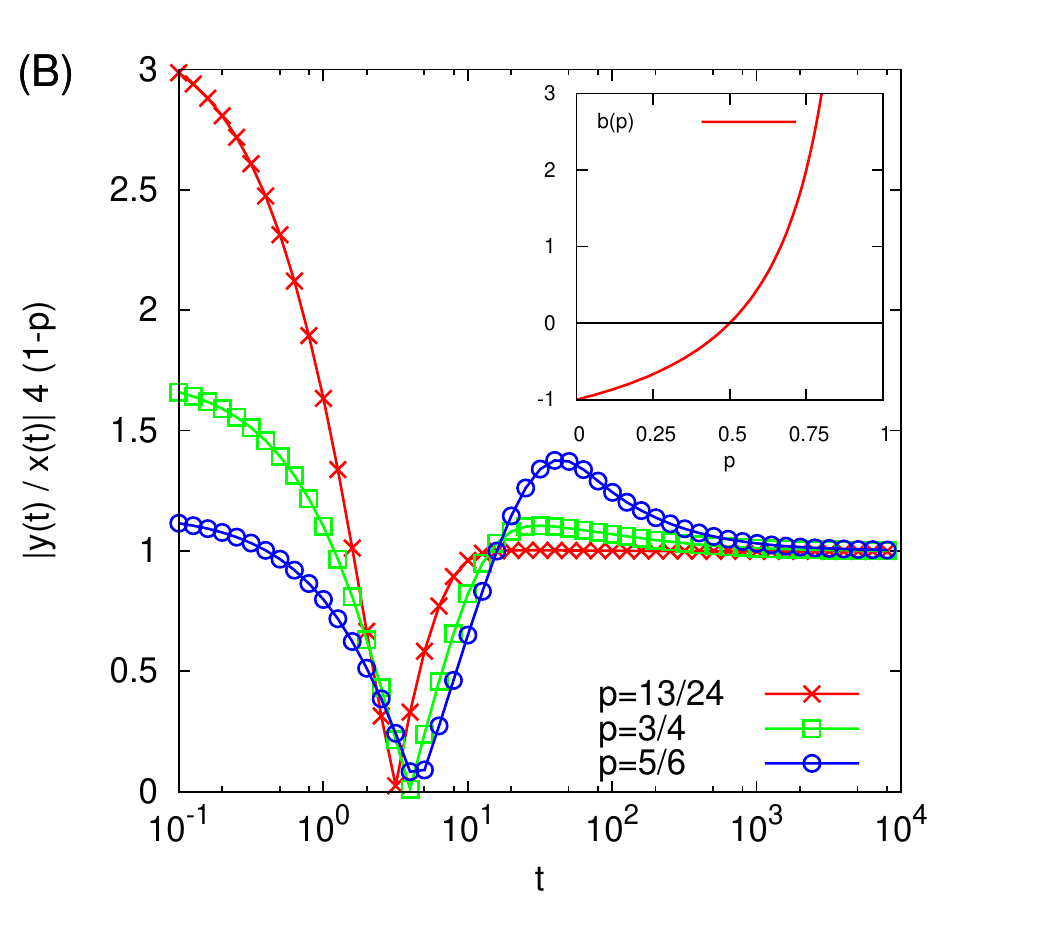}
\label{fig-closureVaryp-ratio}}
\caption{\label{fig-closureVaryp} (Color online) Numerical
investigation of the behaviour of the reduced model,
Eqs.~(\ref{eq-reducedXYZ}), near the coexistence fixed point, $P_0$,
for a range of values of $p$ for which the fixed point is
attractive. Panel (A) shows $\widetilde{X}(t)$ for $p$ taking the
values $\frac{13}{24}$(crosses), $\frac{3}{4}$(squares) and
$\frac{5}{6}$(circles). The data have been compensated by the
$t$-scaling predicted by Eq.~(\ref{eq-predX}) with $b(p)$ given by
Eq.~(\ref{eq-bValue}). Panel (B) shows that the ratios
$Y(t)/\widetilde{X}(t)$ are in agreement with Eq.~(\ref{eq-predRatio})
for each value of $p$. The inset graphs the dependence of $b(p)$ on
$p$ expressed in  Eq.~(\ref{eq-bValue}).} 
\end{center}
\end{figure*}

\begin{figure*}
\begin{center}
\subfigure{
\includegraphics[width=0.9\columnwidth]{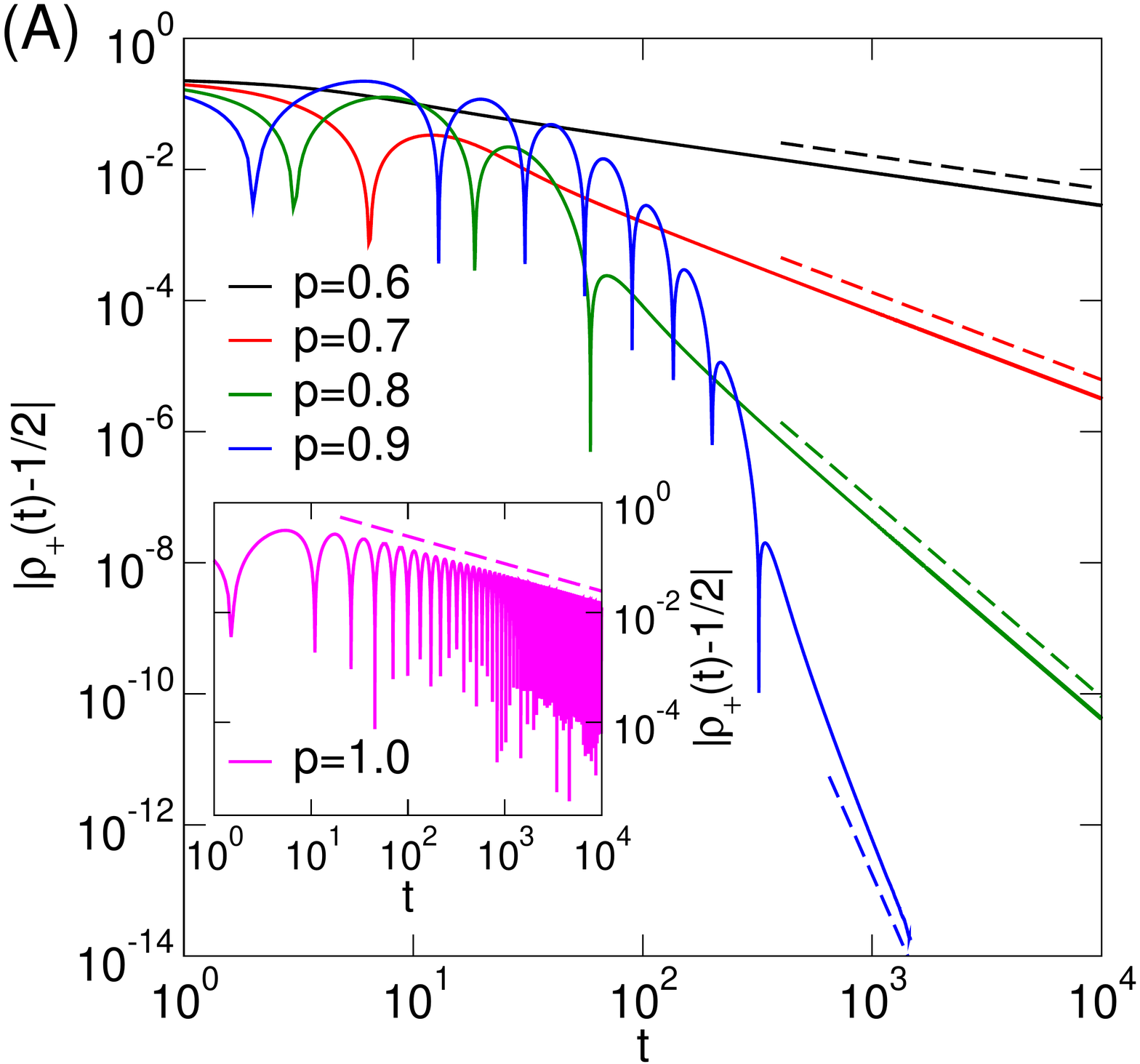} 
\label{fig-powerlaw-1}}
\subfigure{
\includegraphics[width=0.9\columnwidth]{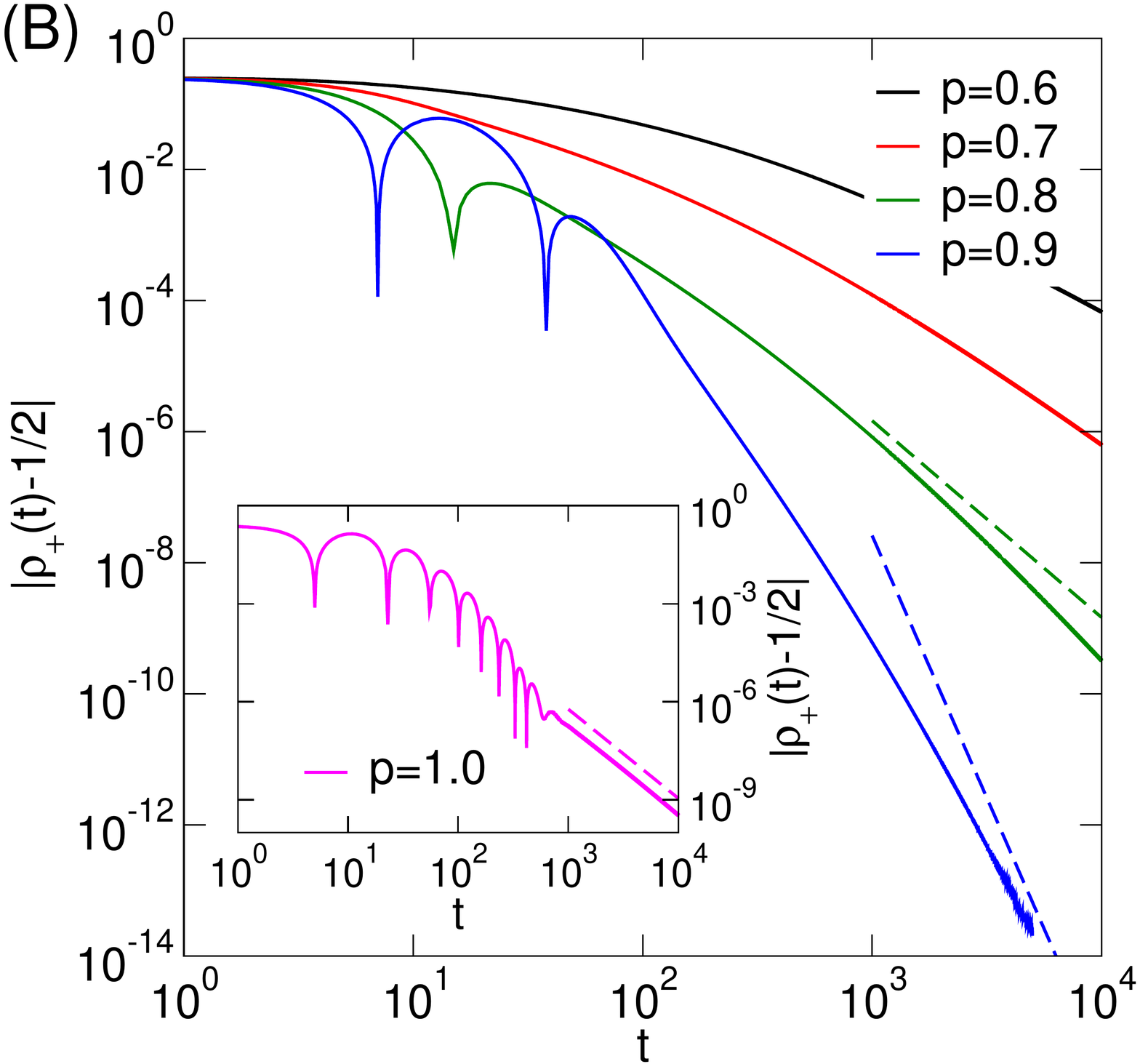}
\label{fig-powerlaw-2}}
\caption{\label{fig-powerlaw} (Color online) Convergence of $\rho_+$ to the
coexistence state $\rho_+=1/2$ for $p>1/2$. (A) Reduced model,
Eqs.~(\ref{eq-reduced})--(\ref{eq-R+}).  The final approach to coexistence
is a power law with an exponent given approximately by the theoretical
expression $b(p)=(2p-1)/(p-1)$ (dashed lines).  Inset: the decay for $p=1.0$
seems to be oscillatory for all times, with an amplitud that decays as
$t^{-0.5}$ (dashed line).  (B) Full MF
equations~(\ref{eq-MF}).  The approach to coexistence is
approximately power law.  Dashed lines indicate theoretical exponents
$b(p)$. Inset: the $p=1.0$ case is special, where the decay seems to
be a pure power law with exponent close to $2.73$ (dashed line).}
\end{center}
\end{figure*}

Replacing $\rho_-=1-\rho_+$ throughout, it is convenient to rewrite 
Eqs.~(\ref{eq-reduced})--(\ref{eq-R+}) in terms of the three
variables $(X(t), Y(t), Z(t))$ defined as
\begin{eqnarray}
\nonumber X&=&\rho_+\\
\label{eq-reducedVariables1}Y &=& \mu_+\rho_- - \mu_-\rho_+\\
\nonumber Z &=& (\mu_+\rho_- + \mu_-\rho_+)^{-1}.
\end{eqnarray}
In terms of these variables, we have the system
\begin{subequations}
\begin{alignat}{3}
\label{eq-reducedModel2X} \dd{X}{t} &= \frac{2 (2p-1)}{1-Y^2Z^2}\,X
(1-X)YZ\left(1-Z\left|Y\right|\right)\\ 
\label{eq-reducedModel2Y} \dd{Y}{t} &= \frac{1}{2}X(1-X)(1-2X)
-(1-p)Y \\ \nonumber &+\frac{2p-1}{2(1-Y^2Z^2)}\,Z F_Y(X,Y,Z)\\
\label{eq-reducedModel2Z} \dd{Z}{t} &=
\frac{Z^2 \left[F_Z(X,Y,Z)-X(1-X) \right]}{2(1-Y^2Z^2)},
\end{alignat}
\label{eq-reducedXYZ}
\end{subequations}
where $F_Y(X,Y,Z)$ and $F_Z(X,Y,Z)$ are complicated multivariate
polynomials which are written out at the end of this section
[Eqs.~(\ref{FY}) and (\ref{FZ})]. The advantage of 
this system is that it is obvious that it has a new fixed point,
\begin{displaymath}
P_0 : (X,Y,Z) = \left(\frac{1}{2},0,0\right),
\end{displaymath}
which corresponds to the self-similar solution of the full MF
equations, Eqs.~(\ref{eq-MF}), found in Sec.~\ref{sec-MF}. We refer to 
$P_0$ as the coexistence fixed point since it describes the situation
in which both populations have size one half. Note that the consensus
fixed points, $P_-$ and $P_+$ are both mapped to $Z=\infty$ in these
variables. Using the system of Eqs.~(\ref{eq-reducedXYZ}) we can try
to probe the stability of the coexistence fixed point. The dynamical
system given by Eqs.~(\ref{eq-reducedXYZ}) cannot be
linearized about $P_0$. Notice, for example, that the lowest power of
$Z$ in the third equation is $Z^2$ so there is no linearization around
$Z=0$. Standard methods of linear stability analysis are therefore
not applicable here. Instead, let us shift the $X$ variable, $X
= \widetilde{X}+\frac{1}{2}$, and look for a scaling solution near
$P_0$:
\begin{eqnarray*}
\widetilde{X}(t) &\sim& X_0\,t^{-a}\\
Y(t) &\sim& Y_0\,t^{-b}\\ Z(t) &\sim& Z_0\,t^{-c}.
\end{eqnarray*}
The powers $a$, $b$ and $c$ must be all positive if the coexistence
fixed point is to be attractive as $t\to\infty$. Some trial and error
is required to identify the leading order terms on the RHS of
Eqs.~(\ref{eq-reducedXYZ}) due to the
large number of terms.  However, this work is greatly simplified when
we note that all terms that appear in the functions $F_Y$ and
$F_Z$ [Eqs.~(\ref{FY}) and (\ref{FZ})] are subleading in the
neighbourhood of the coexistence fixed point.  Then, 
the leading terms on the two sides of Eq.~(\ref{eq-reducedModel2Z}) are
\begin{displaymath}
-c\,Z_0\,t^{-c-1} \sim -\frac{1}{8}\,Z_0^2\,t^{-2c},
\end{displaymath}
so that
\begin{equation}
\label{eq-cRule}
c=1 \hspace{0.25cm}\mbox{and}\hspace{0.25cm} Z_0=8.
\end{equation}
With $c=1$, the leading terms on the two sides of
Eq.~(\ref{eq-reducedModel2X}) are
\begin{displaymath}
-a \, X_0 \, t^{-a-1} \sim \frac{1}{2}\,(2\,p-1)\,Y_0\,Z_0\,t^{-b-c} =
 4\,(2\,p-1)\,Y_0\,t^{-b-1},
\end{displaymath}
which leads us to conclude that
\begin{equation}
\label{eq-aRule}
a=b \hspace{0.25cm}\mbox{and}\hspace{0.25cm} \frac{X_0}{Y_0} =
-\frac{4\,(2\,p-1)}{b}.
\end{equation}
Finally, with $c=1$ and $a=b$, the leading terms in
Eq.~(\ref{eq-reducedModel2Y}) are
\begin{displaymath}
-b\,Y_0\,t^{-b-1} \sim \left( \frac{-X_0}{4}+(p-1)\,Y_0\right) t^{-b}.
\end{displaymath}
This is impossible unless the coefficient of $t^{-b}$ vanishes on the
RHS of Eq.~(\ref{eq-reducedModel2Y}) (there is a subleading  term of
order $t^{-b-1}$ which could then balance the LHS). Therefore we must
have
\begin{equation}
\label{eq-bRule}
\frac{X_0}{Y_0} = 4\,(p-1).
\end{equation}
Combining Eqs.~(\ref{eq-aRule}) and (\ref{eq-bRule}) we find
\begin{equation}
\label{eq-bValue}
b = -\frac{2\,p-1}{p-1}.
\end{equation}
Thus all three exponents are determined along with the amplitude
$Z_0$. The amplitudes $X_0$ and $Y_0$ are arbitrary but their ratio is
fixed and given by Eq.~(\ref{eq-bRule}). To summarise, the reduced
model predicts the following behaviour near the coexistence fixed
point:
\begin{eqnarray}
\label{eq-predX}\widetilde{X}(t)  &\sim& X_0\,t^{-b(p)}\\
\label{eq-predY}Y(t) &\sim& Y_0\,t^{-b(p)}\\
\label{eq-predZ}Z(t) &\sim& 8\,t^{-1}\\
\label{eq-predRatio}\frac{X_0}{Y_0} &=& 4\,(p-1),
\end{eqnarray}
with $b(p)$ given by Eq.~(\ref{eq-bValue}).  These predictions are
validated against numerical solutions of Eqs.~(\ref{eq-reducedXYZ}) in
Figs.~\ref{fig-closureVaryIC} and \ref{fig-closureVaryp}.  We find
that $b>0$ for $\frac{1}{2} < p \leq 1$ [see the inset of
Fig.~\ref{fig-closureVaryp-ratio}]. The coexistence fixed point is
therefore attractive for values of $p$ in this range and repulsive
otherwise.  An interesting observation is that the approach of $X$ to
the fixed point $X=1/2$ shows damped oscillations for $p \gtrapprox 0.66$, while
for $p \lessapprox 0.66$ the approach is monotonic, as we can see in
Fig.~\ref{fig-powerlaw-1}.   This transition between the monotonic and
the oscillatory regimes is also found in the full MF model 
[Eqs.~(\ref{eq-MF})] at a value $p_o \simeq 0.8$ [see
Fig.~\ref{fig-powerlaw-2}].  In Fig.~\ref{fig-powerlaw-2} we also compare
the theoretical decay  $t^{-b(p)}$ from Eq.~(\ref{eq-bValue}) (dashed
lines) with the one from the MF Eqs.~(\ref{eq-MF}) (solid
curves), for three different values of $p$.  We observe that the
agreement improves as $p$ gets larger.

Below we provide for completeness the explicit formulae for the
multivariate polynomials appearing  on the right hand side of
Eqs.~(\ref{eq-reducedXYZ}) defining the reduced model. 

\begin{widetext}
\begin{eqnarray}
\label{FY}
F_Y(X,Y,Z) &=&-2X^5 + 5 X^4 -4 X^3 - 2X^2 Y + X^2 - 4 XY^2 + 2XY + 2
Y^3 Z + 2Y^2+\left(4XY^2Z - 2Y^2Z + 2X^2YZ\right. \nonumber \\
&&\left.-2XYZ-2Y+2X^5 Z -5X^4Z+4X^3Z -
X^2Z\right)\,\left|Y\right|. \\
\label{FZ}
F_Z(X,Y,Z) &=& -X^2Y^2Z^2 + 2XY^3Z^2-2XY- Y^3Z^2+Y + (2
p-1)\left[2 XY^3 Z^2 -Y^3Z^2 + 2Y^2Z -4X^3YZ \right. \nonumber \\ & & +6X^2YZ -2
XYZ +10XY -5Y -X^4Z +2X^3Z -X^2Z + \left(X^4Z^2 + 4X^3YZ^2 -2X^3Z^2 -
6X^2YZ^2\right. \nonumber \\ & & \left.\left.+X^2Z^2 + "XYZ^2 -12 XYZ -2 Y^2Z^2
+6YZ\right)\,\left|Y \right|\right].
\end{eqnarray}
\end{widetext}

\section{Conclusions}
\label{sec-conclusions}

To conclude, we have studied a variant of the voter model in which
each agent is endowed with a fitness parameter, $k$, in addition to
its opinion variable.  Agents interact by pairs, and a single parameter $p$  determines the probability that the agent with the higher $k$--value wins.  When an agent wins an interaction, its $k$--value is increased by $1$, and the looser agent changes opinion.  The distribution of $k$--values in the
population therefore co-evolves with the opinion dynamics.  The rates of opinion change therefore depend on the past history of the agents.  Our model has aspects in common with several models which have been studied in the
literature, particularly the competitive population dynamics studied
in \cite{ben-naim_structure_2006}, the Partisan Voter
Model \cite{masuda_heterogeneous_2010} and the non-Markovian voter
model studied in \cite{stark_decelerating_2008}.  Through a
combination of numerical simulations and analysis we showed that there
is a coexistence state in which both populations have a size similar
to $N/2$, and their mean $k$-value increases linearly in time.  This
coexistence state is attractive on average when $p>1/2$ and repulsive
on average when $p<1/2$. As a consequence, the consensus time is
increased relative to the standard voter model when $p>1/2$, whereas
the system is driven to fast consensus when $p<1/2$.  The dynamics in
the $p>1/2$ case exhibits interesting properties, including a
monotonic approach to the coexistence state, as well as damped
oscillations that decay as a power law in time with a non-universal
exponent.  A quantitative explanation of these stability properties
was provided in the context of a reduced 3--D dynamical system based on the full rate equations of the model.  One of the outstanding mysteries at this point is that there is little real indication
in any of our analysis of why the model with $p=1$ has a larger scaling exponent $\beta \simeq 1.45$ for the consensus time, compared to the exponent $\beta \simeq 1.0$ for $1/2<p<1$ (see Sec.~\ref{sec-consensusTimes}). This is a topic for future investigation.  Another obvious avenue for further investigation would be to study the spatial coarsening properties of the model on regular lattices, since we already know that for $p=1/2$ the coarsening dynamics is equivalent to those in the regular voter model.

\section*{Acknowledgments}
This research was supported in part by the National
Science Foundation under Grant No. NSF PHY11-25915
and the EPSRC under grants No. EP/M003620/1 and
EP/E501311/1. CC is grateful for the hospitality Kavli
Institute for Theoretical Physics where the first draft of
the manuscript was written.  FV acknowledges financial support from
CONICET (PIP 0443/2014).  
\bibliography{references}
\end{document}